\documentclass[11pt,english]{article}
\usepackage[T1]{fontenc}
\usepackage[latin1]{inputenc}
\usepackage{babel}
\usepackage{graphics}

\makeatletter

\providecommand{\LyX}{L\kern-.1667em\lower.25em\hbox{Y}\kern-.125emX\@}
\let\SF@@footnote\footnote
\def\footnote{\ifx\protect\@typeset@protect
    \expandafter\SF@@footnote
  \else
    \expandafter\SF@gobble@opt
  \fi
}
\expandafter\def\csname SF@gobble@opt \endcsname{\@ifnextchar[
  \SF@gobble@twobracket
  \@gobble
}
\edef\SF@gobble@opt{\noexpand\protect
  \expandafter\noexpand\csname SF@gobble@opt \endcsname}
\def\SF@gobble@twobracket[#1]#2{}

\usepackage[T1]{fontenc}
\usepackage[latin1]{inputenc}
\usepackage{graphics}

\makeatletter

\usepackage[T1]{fontenc}
\usepackage[latin1]{inputenc}
\usepackage{graphics}

\makeatletter

\usepackage[T1]{fontenc}
\usepackage[latin1]{inputenc}
\usepackage{graphics}

\makeatletter

\usepackage[T1]{fontenc}
\usepackage[latin1]{inputenc}
\usepackage{graphics}

\makeatletter

\usepackage{amsmath}
\usepackage{amssymb}
\usepackage{epsfig}
\usepackage{psfrag}
\usepackage{graphicx}
\usepackage{here}

\normalsize

\makeatother

\makeatother

\makeatother

\makeatother

\makeatother
\begin{document}

\title{Emergence of Jupiter's vortices and jets from random initial conditions\\
}

\author{F. BOUCHET{\small \( ^{1,2} \)} and T. DUMONT{\small \( ^{3} \)}\\
\\
{\small \( ^{1} \)Institut Fourier, UMR 5582, Grenoble France}\\
{\small \( ^{2} \)Dipartimento di Energetica {}``Sergio Stecco'',
Universit\`a degli studi di Firenze, Italy}\\
{\small \( ^{2} \)Laboratoire d'Analyse Numérique-CNRS UMR 5585 Université
Lyon 1 France}\\
{\small \nonumber }}

\maketitle
Received:

Author version: \date

Author contact : bouchet@dma.unifi.it 

\newpage

\begin{abstract}
We explain the emergence and stability of the most important jets
and vortices, in the highly turbulent Jupiter's atmosphere, by a statistical
mechanics of the potential vorticity mixing. Using the Quasi-Geostrophic
1-1/2 layer, with topography, when the Rossby deformation radius is
small, we predict strong jets. These jets can be either zonal, or
closed into a ring structure like the Great Red Spot one. We reproduce
the GRS observed velocity field to a very good quantitative accuracy.
For smaller vortices, or for stronger topography curvature, we reproduce
the characteristics properties of the White Ovals or of the cyclonic
Brown Barges. The link between their shape, topography and surrounding
shear is explicitly described. We obtain very strong qualitative results
for the Jupiter's vortices. For instance, any of these vortices must
be on topography extrema (in the reference frame moving with the structure),
the shear in the active layer is larger than the shear in the deep
layer. On a same latitudinal band, the velocity of the vortex is related
to their latitude. These theoretical predictions are in accordance
with the observed properties of Jovian vortices. 
\end{abstract}

\section{Introduction}

Atmospheric and oceanic flows have the property to organize into large
scale jets or vortices. Due to the large difference between typical
forcing and inertial time scales, this organization is remarkably
stationary in the case of Jupiter's troposphere. The understanding
of the stability and of the detailed structure of these flows is thus
render easier than for any other geophysical flows. Moreover, the
excellent quality of the data obtained from space probes, makes easy
a precise comparison of theoretical predictions with actual flows. 

\begin{figure}
{\centering \resizebox*{1.1\textwidth}{!}{\includegraphics{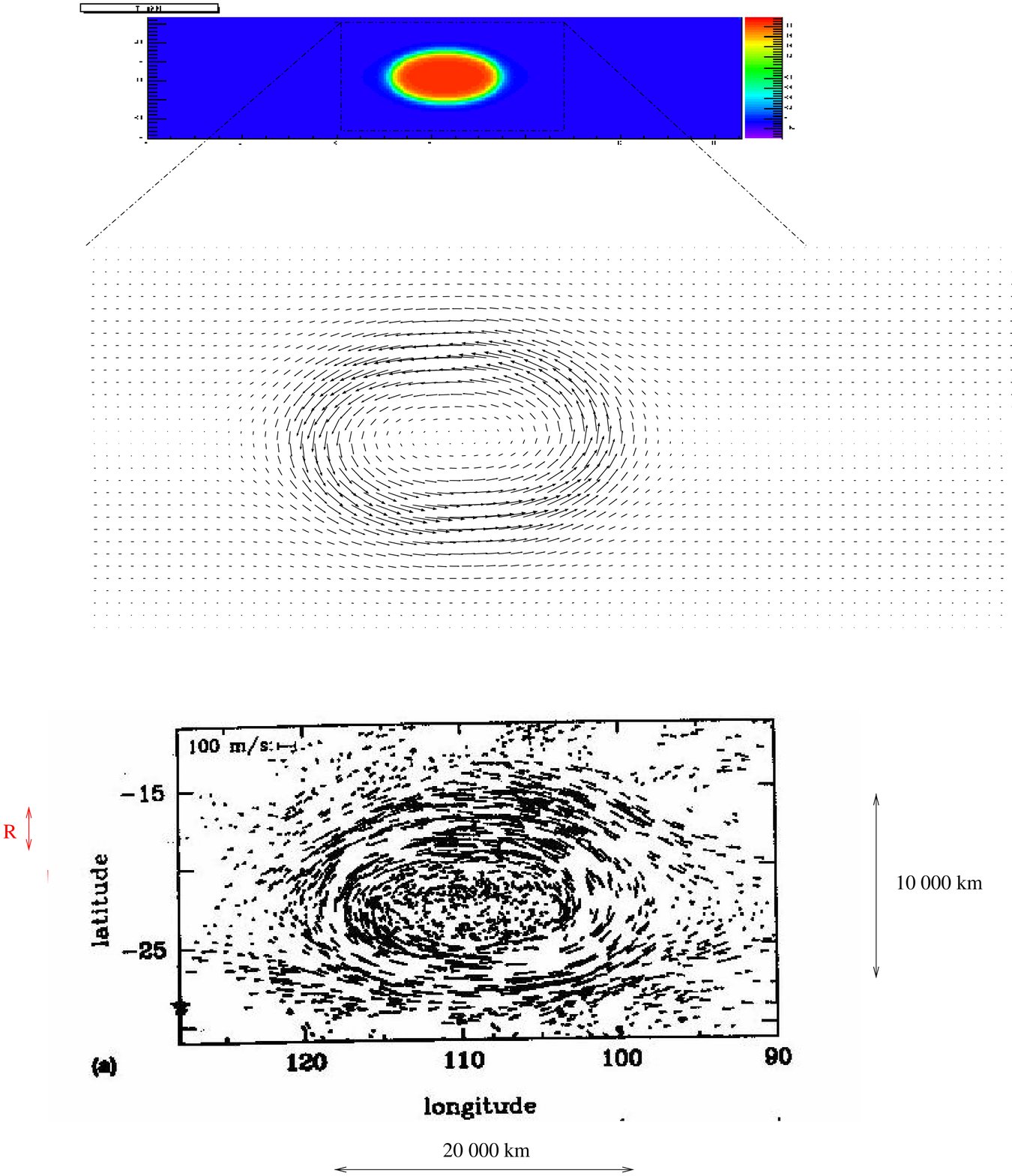}} \par}

\caption{Upper part: vorticity field and velocity field for the statistical
equilibrium modeling the Great Red Spot. Lower : the observed velocity
field, from Dowling and Ingersoll (1988). The actual values of the
jet maximum velocity, jet width, vortex width and length fit with
the observed ones. The strong jet is the interface between two phases,
each corresponding to different Potential Vorticity mixing. It obeys
a minimal length variational problem, balanced by the effect of the
deep layer shear.\label{fig:Emergence_Numerique_tache_rouge} }
\selectlanguage{english}
\end{figure}

As in the oceans and in the earth atmosphere, these flows are often
organized into narrow jets. They can zonally flow around the planet,
like the eastward jet at 24\( ^{0} \) latitude in the northern hemisphere
of Jupiter (Maxworthy 1984), or alternatively organize into rings,
forming vortices, like the rings shed by the meandering of the Gulf-Stream
in the western Atlantic Ocean. The flow field in Jupiter most famous
feature, the Great Red Spot, is an oval-shaped jet, rotating in the
anticyclonic direction and surrounding an interior area with a weak
mean flow (Dowling and Ingersoll 1989) (see figure \ref{fig:Emergence_Numerique_tache_rouge}).
Robust cyclonic vortices are also observed with a similar jet structure
(Hatzes et al 1981), see figure \ref{fig:Emergence_Numerique_Jet_Tache}.
Smaller features, as the White Ovals, have also an oval shape but
without the jet ring structure. For a recent review on the dynamics
of Jupiter's atmosphere, see Ingersoll and collaborators (2003). 

A number of numerical studies have been led to model the Jupiter's
vortices (see Ingersoll and collaborators 2003, for a review). The
Kida vortex (Kida 1981) has been used to explain the typical oval
shape of such vortices (Polvani and collaborators 1990). Some soliton
like structures have been also describe (Petviashvili 1981, ????)
with similar shapes. Anyway, none of these analytical and theoretical
models, reproduce both the oval shape and ring structure of the Great
Red Spot. We will argue that the strong jets are the consequence of
a small value of the Rossby deformation radius, for very energetic
flows. With such parameters, the effect of a topography (deep flow
and beta effect) will be necessary to explain the typical oval shape.
Moreover, such jets and vortices are in a turbulent surrounding, and
the persistence of their strength and concentration in the presence
of eddy mixing is intriguing and should be explained.\\
The explanation proposed in this paper is based on a statistical mechanical
approach: the narrow jet or vortex appears as the most probable state
of the flow after a turbulent mixing of potential vorticity, taking
into account constraints due to the dynamical conserved quantities,
especially energy. Such a statistical theory has been first proposed
for the two-dimensional incompressible Euler equations by Kuz'min
(1982), Robert (1990), Robert and Sommeria (1991), Miller (1990),
see Sommeria (2001) for a recent review. This theory predicts an organization
of 2D turbulence into a steady flow, superposed with fine scale, 'microscopic'
vorticity fluctuations. This is by far the most likely result of random
stirring, so the evolution to this statistical equilibrium is in practice
irreversible. Complete vorticity mixing is prevented by the conservation
of the energy, which can be expressed as a constraint in the accessible
vorticity fields. A similar, but quantitatively different, organization
had been previously obtained with statistical mechanics of singular
point vortices, instead of continuous vorticity fields (Onsager 1949,
Joyce and Montgomery 1973). The possibility of using such ideas to
explain the Great Red Spot has been explicitly quoted since the first
works on the 2D Euler statistical mechanics by Robert (1990), Miller
(1990), Sommeria et al (1991), Miller Weichman and Cross (1992), Turkington,
Majda and DiBattista (2001), but without explicit predictions. 

In the study of geophysical flows, the Rossby deformation radius is
a central parameter, as it defines a typical scale for the variation
of the pressure. In a previous paper (Bouchet and Sommeria, 2002),
we have analytically described the statistical equilibrium states,
for the Quasi-Geostrophic equation, in the limit of a small Rossby
deformation radius. In this limit, the equilibrium flows are characterized
by strong jets, either zonal and flowing around the planet or forming
closed vortices, depending on the parameters. When a topography is
considered, these vortices have the typical shape of Jovian ones'.
We have also shown that these equilibrium are able to reproduce quantitatively
all the main characteristics of the Great Red Spot. In section \ref{sec:Equilibrium_Small_Rossby},
we will give a simplified derivation of these results. Whereas in
Bouchet and Sommeria 2002, the statistical equilibrium were analyzed
for the simple case of a potential vorticity distribution made of
two levels, we extensively discuss the generalization of the results
to any PV distribution. We show that the main results do not depend
on the detailed distribution. This is true as a consequence of the
small value of the Rossby deformation radius.

In section \ref{sec_Application_Jupiter}, we discuss the application
of such results to the Jovian vortices. We discuss a simple model
for the Great Red Spot, also present in Bouchet and Sommeria (2002).
We discuss important qualitative predictions for such vortices : they
are located on extrema of the topography, their energy must be greater
than a critical one, the shear on the active layer has to be stronger
than in the deep layer, their typical width is given by an alternative
of the Rhine's scale, which no does not depends on the beta-effect,
but on the topography curvature.

The limit of small Rossby radius is no more valid for smaller vortices,
such as the Brown Barges or the White Ovals. In order to model these
features, we will numerically compute the velocity fields of the statistical
equilibrium states. Using the qualitative comprehension obtained from
the analytical analysis, we will be able to reproduce the main properties
of these flows, either from the destabilization of jets, or from random
Potential Vorticity initial conditions. In section \ref{sec:Emergence_Num_OBC}
we obtain a velocity field close to the White-Oval one. In section
\ref{sec:Emergence_Num_GRS}, we obtain numerically the velocity field
of the Great Red Spot which is accurately compared to the observed
one's. In section \ref{sec_Brown_Barges}, we obtain the peculiar
velocity field of the Brown Barges, with a jet like structure in the
meridional direction, and a strong shear in the zonal one.

In section \ref{sec:Emergence_Num_Jets}, we comment results of Ellis,
Haven and Turkington (2002), on the stability of statistical equilibrium.
We illustrate the corresponding results by numerical experiment of
stabilization or destabilization of strong jets. This complete the
explanation of the emergence and of the stability of Jupiter's features
in the Jupiter's turbulent atmosphere.

\section{Statistical mechanics of the Quasi-Geostrophic equation in the limit
of small Rossby deformation radius}

In this section we present the Quasi-Geostrophic 1-1/2 layer model
and the main ideas of the potential vorticity statistical mechanics.
This theory describes the most probable state, emerging from a random
PV field with a given PV distribution and energy. The main hypothesis
is that these equilibrium structures emerge from the very complex
dynamical mixing. These stationary states have been described analytically
in a previous work (Bouchet and Sommeria 2002), for the Quasi-Geostrophic
model, in the limit of small Rossby deformation radius. In this section
we sketch the main ideas of this derivation and the main results and
consequence in the context of the Jovian troposphere. These results
explain in particular the formation of jets or vortices from random
initial conditions. Such vortices have the annular jet structure characteristic
of the Great Red Spot and their main characteristics are analytically
related. 

In the following sections we will propose numerical simulation, illustrating
these main results, and permitting to compare them to the main structures
of the Jovian troposphere : the strong jets, the north hemisphere
Brown Barges, the White Ovals, and the Great Red Spot.

\subsection{The dynamical system\label{secdyn}}

We start from the barotropic 1-1/2 Quasi Geostrophic (QG) equation
: \begin{equation}
\label{QG}
\frac{\partial q}{\partial t}+{\textbf {v}\cdot \nabla }q=0
\end{equation}
\begin{equation}
\label{dir}
q=-\Delta \psi +\frac{\psi }{R^{2}}-Rh(y)
\end{equation}
\begin{equation}
\label{u}
{\textbf {v}}=-{\textbf {e}_{z}}\wedge \nabla \psi 
\end{equation}
 where \( q \) is the potential vorticity (PV), advected by the non-divergent
velocity \( {\textbf {v}} \), \( \psi  \) is the stream function%
\footnote{\label{notepsi}We choose for the stream function \( \psi  \) the
standard sign convention used for the Euler equation, which is just
the opposite as the one commonly used in geophysical fluid dynamics.
Our stream function \( \psi  \) is therefore proportional to the
opposite of the pressure fluctuation in the northern hemisphere and
to the pressure fluctuation in the southern hemisphere, as the planetary
vorticity sign is reversed. The signs of \( q \) and \( {\textbf {v}} \)
are not influenced by this choice of sign for \( \psi  \). 
}, \( R \) is the internal Rossby deformation radius between the layer
of fluid under consideration and a deep thicker layer, unaffected
by the dynamics. \( x \) and \( y \) are respectively the zonal
and meridional coordinates (\( x \) is directed eastward and \( y \)
northward). 

The term \( Rh(y) \) represents the combined effect of the planetary
vorticity gradient and of a given stationary zonal flow in the deep
layer, with stream function \( \psi _{d}(y) \): \( Rh(y)=-\beta y+\psi _{d}/R^{2} \).
This deep flow induces a constant deformation of the free surface,
acting like a topography on the active layer%
\footnote{\label{noteh}A real topography \( \eta (y) \) would correspond to
\( Rh(y)=-f_{0}\eta (y)/h_{0} \) where \( f_{0} \) is the reference
planetary vorticity at the latitude under consideration and \( h_{0} \)
is the mean upper layer thickness. Due to the sign of \( f_{0} \),
the signs of \( h \) and \( \eta  \) would be the same in the south
hemisphere and opposite in the north hemisphere. As we will discuss
extensively the Jovian south hemisphere vortices, we have chosen this
sign convention for \( h \). 
}. We shall therefore call \( h(y) \) the 'topography'. We scale the
topography with the Rossby deformation radius \( R \). This particular
choice will be of importance in the study of the limit \( R\rightarrow 0 \)
(section \ref{sec:Equilibrium_Small_Rossby}) and we will show that
this scaling is the appropriate one to study Jovian vortices. 

We define the QG equations (\ref{QG},\ref{dir}) with periodic boundary
conditions (\( 4\pi  \) periodic in the zonal direction and \( \pi  \)
periodic in the meridional one for all numerical computations of this
article). The analytical study in Bouchet and Sommeria 2002 has shown
that, in the channel geometry, due to the small value of the Rossby
deformation radius, the equilibrium organization of the flow is local.
For instance vortices are located on topography extrema and their
structure and shape is determined by the topography curvature and
is independent on boundary conditions. In the periodic boundary conditions
case, we will show in the following that this is still true. This
local organization explains why periodic boundary condition is well
suited to vortices structure studies. On the contrary, the global
organization of zones and bands on the planet scale should be tackled
using a real spherical geometry, for instance in the Shallow Water
model. This more general problem will not be considered in this article. 

We model one zone and band area by a periodic topography: \begin{equation}
\label{Topographie}
h(y)=-2a\cos (2y)
\end{equation}
 As the relevant scale is defined by the latitudinal variations of
the topography, we do not respect the actual zonal band aspect ratio,
and we scale the domain size on the latitudinal zone-band extension.
In our dimensionless variable \( R=\pi R^{\star }/L^{\star } \) where
\( R^{\star } \) is the actual internal Rossby deformation radius
and \( L^{\star } \) is the latitudinal extension of the zone-band
domain. \\

Let \( \langle f\rangle \equiv \int _{D}fd^{2}{\textbf {r}} \) be
the average of \( f \) on \( D \) for any function \( f \). Physically,
as the stream function \( \psi  \) is related to the geostrophic
pressure, \( \langle \psi \rangle  \) is proportional to the mean
height at the interface between the fluid layer and the bottom layer,
and due to the mass conservation it must be constant (Pedlosky 1987).
We make the choices \( \langle \psi \rangle =0 \) and \( \langle h\rangle =0 \)
without loss of generality. The total circulation is \( \langle q\rangle =\langle -\Delta \psi +\psi /R^{2}-Rh(y)\rangle  \)
is equal \( \langle \psi /R^{2}\rangle  \) due to the periodic boundary
conditions. Therefore \( \langle q\rangle =0 \). 

Due to the periodic conditions for \( \psi  \), the linear momentum
is also equal to 0, \begin{equation}
\label{vitessemoy}
\langle {\textbf {v}}\rangle =0
\end{equation}

The energy \begin{equation}
\label{ene}
E={\frac{1}{2}}\int _{D}\left( q+Rh\right) \psi d^{2}{\textbf {r}}={\frac{1}{2}}\int _{D}\left[ (\nabla \psi )^{2}+\frac{\psi ^{2}}{R^{2}}\right] d^{2}{\textbf {r}}
\end{equation}
 is conserved (we note that the first term in the right hand side
of (\ref{ene}) is the kinetic energy whereas the second one is the
gravitational available potential energy).

The Casimir integrals \begin{equation}
\label{cas}
C_{f}(q)=\int _{D}f(q)d^{2}{\textbf {r}}
\end{equation}
 for any continuous function \( f \), in particular the different
moments of the PV, are also conserved.

\subsection{The statistical mechanics on a two PV levels configuration.}

\subsubsection{The macroscopic description.}

The QG equations (\ref{QG}) (\ref{dir}) are known to develop very
complex vorticity filaments. Because of the rapidly increasing amount
of information it would require, a deterministic description of the
flow for long time is both unrealistic and meaningless. The statistical
theory adopts a probabilistic description for the vorticity field.
We consider the local probability to have some PV at some points.
The statistical equilibrium is then the most probable state for a
random PV field with fixed dynamical invariants. 

The statistical equilibrium therefore depends on the energy (\ref{ene})
and on the infinite number of Casimirs (\ref{cas}) (PV distribution).
For pedagogical reasons, we will consider the most simple case we
will suppose a distribution made of two PV levels, denoted \( q=a_{1} \)
and \( q=a_{-1} \). The results may however be generalized (Robert
and Sommeria 1992). In section \ref{sec:Equilibrium_Small_Rossby},
we will explain why the study of the equilibrium structures is independent
of the actual PV distribution, at the lower order when the Rossby
radius goes to zero.

The two values of the PV \( q=a_{1} \) and \( q=a_{-1} \), and the
areas \( A \) and \( (1-A) \) they respectively occupy in \( D \),
will be conserved by the inertial dynamics (this is then equivalent
to the conservation of all the Casimirs (\ref{cas})). The determination
of the statistical equilibrium then depends only on the energy \( E \),
on the two PV levels \( a_{1} \) and \( a_{-1} \) and on the area
\( A \). The number of free parameters can be further reduced by
appropriate scaling. Indeed a change in the time unit permits to define
the PV levels up to a multiplicative constant. We choose for the sake
of simplicity : \begin{equation}
\label{csttemps}
\frac{a_{1}-a_{-1}}{2}=1
\end{equation}
 and define the non-dimensional parameter \( B \) as : \begin{equation}
\label{B}
B\equiv \frac{a_{1}+a_{-1}}{2}
\end{equation}
 As discussed previously the mean PV is equal to zero, this imposes
that \( a_{1}A+a_{-1}(1-A)=0 \). This means that \( a_{1} \) and
\( a_{-1} \) must be of opposite sign and, using (\ref{csttemps})
and (\ref{B}), \( A=(1-B)/2 \). The distribution of PV levels is
therefore fully characterized by the single asymmetry parameter \( B \),
which takes values between -1 and +1. The symmetric case of two PV
patches with equal area \( A=1/2 \) corresponds to \( B=0 \), while
the case of a patch with small area (but high PV, such that \( \left\langle q\right\rangle =0 \))
corresponds to \( B\rightarrow 1 \). Note that we can restrict the
discussion to \( B\geq 1 \) as the QG system is symmetric by a change
of sign of the PV.\\

The two PV levels mix due to turbulent dynamics, and the resulting
state is locally described by the local probability (local area proportion)
\( p({\textbf {r}}) \) to find the first level at the location \( {\textbf {r}} \).
The probability to find the complementary PV level \( a_{-1} \) is
\( 1-p \), and the locally averaged PV at each point is then \begin{equation}
\label{PVmean}
\overline{q}({\textbf {r}})=a_{1}p({\textbf {r}})+a_{-1}(1-p({\textbf {r}}))=2\left( p-\frac{1}{2}\right) +B
\end{equation}
 where the second relation is obtained by using (\ref{csttemps})
and (\ref{B}).

Since the patch with PV level \( a_{1} \) is mixed but globally conserved,
the integral of its density \( p \) over the domain must be equal
to the initial area \( A \), \begin{equation}
\label{PVcons}
A\equiv \frac{1-B}{2}=\int _{D}p({\textbf {r}})d^{2}{\textbf {r}}
\end{equation}

We note that the inertial conservation of the Casimir, is taken into
account in the microscopic description, by the knowledge of the distribution
of the PV. However the coarse-graining (macroscopic description) does
not preserves the value of the Casimirs (\ref{cas}): \( C_{f}(q)\neq C_{f}(\overline{q}) \),
except for the first moment. 

The effect of local PV fluctuations on the stream function is filtered
out by integration of equation \ref{dir} (\( \overline{\psi }= \)
\( \psi  \) and \( \overline{{\textbf {v}}}={\textbf {v}} \)), the
stream function and the velocity field are thus fully determined by
the locally averaged PV \( \overline{q} \) as the solution of \begin{equation}
\label{dirmean}
\overline{q}=-\Delta \psi +\frac{\psi }{R^{2}}-Rh(y)\, \, ;\, \, \psi \, \, \, \, periodic\, \, 
\end{equation}

\[
and\, \, \, \, \, \, {\textbf {v}}=-{\textbf {e}_{z}}\wedge \nabla \psi \]
 Therefore the energy is also expressed in terms of the field \( \overline{q} \)
: \begin{equation}
\label{Econs}
E={\frac{1}{2}}\int _{D}\left[ \, \, (\nabla \psi )^{2}+\frac{\psi ^{2}}{R^{2}}\, \, \right] d^{2}{\textbf {r}}=\frac{1}{2}\int _{D}\left( q+Rh\right) \psi d^{2}{\textbf {r}}
\end{equation}
 From now on we forget the \( q \) over-line for the locally averaged
PV and refer to it as the PV.

The central result of the statistical mechanics of the QG equations
(\ref{QG},\ref{dir}) is that the most probable mixing of the potential
vorticity is given by the maximization of the entropy \begin{equation}
\label{ent}
S=-\int _{D}[\, \, p({\textbf {r}})\ln p({\textbf {r}})+(1-p({\textbf {r}}))\ln (1-p({\textbf {r}}))\, \, ]d^{2}{\textbf {r}}
\end{equation}
 under the constraints of the global PV distribution (\ref{PVcons})
and energy (\ref{Econs}). It can be shown that the microscopic states
satisfying the constraints given by the conservation laws are overwhelmingly
concentrated near the Gibbs state. A good justification of this statement
is obtained by the construction of converging sequences of approximations
of the QG equation (\ref{QG},\ref{dir}), in finite dimensional vector
spaces, for which a Liouville theorem holds. This is a straightforward
translation of the work of Robert (1999) for 2D Euler equations. The
sequence of such Liouville measures has then the desired concentration
properties as (\ref{QG},\ref{dir}) enters in the context considered
in Michel \& Robert (1994 b). More recently; Ellis (1999) also discussed
such large deviation results together with other systems.

Once the most probable state is found, we suppose that it describes
observed flows. The ergodicity of the system would be sufficient to
justify this. But, as in usual statistical mechanics (for instance
for gas) this ergodic property of a system is very unlikely to be
proven for any generic system and could moreover appear to be wrong
in general. A weaker property of mixing is however sufficient to justify
statistical mechanics due to the concentration property stated in
the above paragraph. The Gibbs state is most likely to be reached
even if the available microscopic states are not evenly explored.
In practice, the theory can be validated or invalidated only on the
basis of its success or failure to predict well characterized phenomena.

\subsubsection{The Gibbs states\label{sec_Gibbs_states}}

We want to describe the equilibrium structures (Gibbs states). We
thus seek the maxima of the entropy (\ref{ent}) under the constraints
of the area (\ref{PVcons}) and energy (\ref{Econs}): \begin{equation}
\label{Probleme-Variationnel-Entropie}
\max \left\{ S\, |\, {\rm with}\, E=E_{0}\, {\rm and}\, A=(1-B)/2\right\} 
\end{equation}

\begin{figure}

\caption{The area free energy \protect\( f_{C}\left( \phi \right) \protect \)
specifying the free energy functional (\ref{Modica_Variationnel_EnergieLibre_Final}).
For any value of \protect\( C\protect \) , the function \protect\( f_{C}\left( \phi \right) \protect \)
is even and possess two minima \protect\( \pm u\protect \). This
shows that, at equilibrium, at zeroth order in \protect\( R\protect \),
the Potential Vorticity mixing will be described by two phases, characterized
by these two minima. This plot corresponds to the value \protect\( C=10\protect \).}

\label{fig_U}

\par\centering \resizebox*{12cm}{!}{\includegraphics{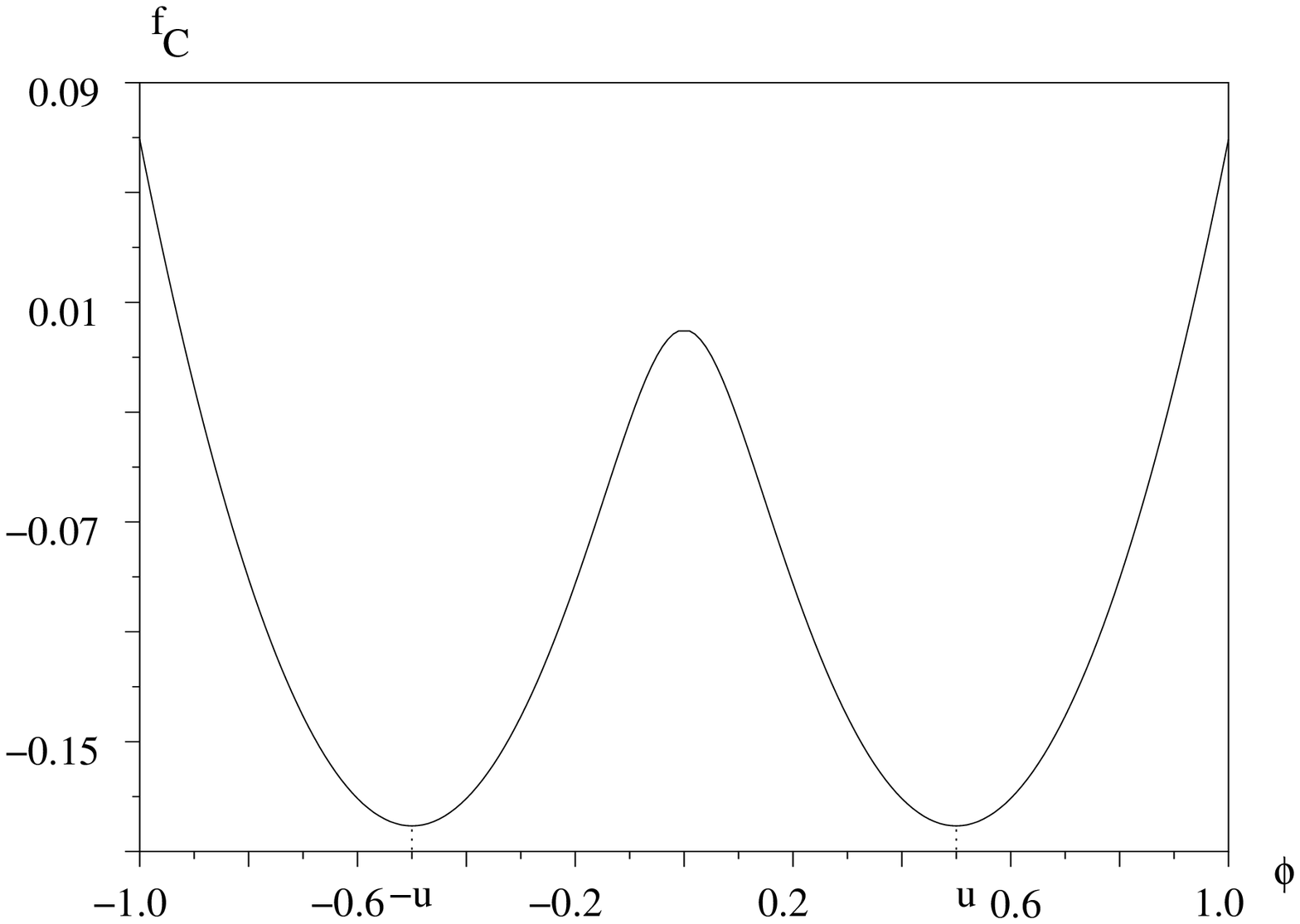}  } \par{}
\end{figure}

In Bouchet and Sommeria (2002), we have studied this variational problem
in the limit of small Rossby deformation radius. The study of such
a variational problem is rendered difficult by the two constraints.
In the following, we will argue that for the present case, this technical
difficulty may be circumvented. We will then proposed a more straightforward
derivation of Bouchet and Sommeria (2002) results. The main ideas
are however the same. For this, let us consider the following variational
problem:

\begin{equation}
\label{Modica_Variationnel_EnergieLibre_Final}
\left\{ \begin{array}{c}
\min \left\{ \mathcal{F}\left[ \phi \right] \, |\, {\rm with}\, \, A\left[ \phi \right] =-\alpha \right\} \\
{\rm with\, \, \, \, }\mathcal{F}=\int _{D}d{\bf r}\, \left[ \frac{R^{2}\left( \nabla \phi \right) ^{2}}{2}+f_{C}(\phi )-R\phi h(y)\right] \, \, \, \, {\rm ,}\, \, \, \, \mathcal{A}\left[ \phi \right] =\int _{D}d{\bf r}\, \phi \\
{\rm and}\, \, \, \, f_{C}\left( \phi \right) =\phi ^{2}/2-\ln \left( \cosh \left( C\phi \right) \right) /C
\end{array}\right. 
\end{equation}
\foreignlanguage{english}{We will call \( \mathcal{F} \) the modified
free energy. This variational problem (\ref{Modica_Variationnel_EnergieLibre_Final})
involves only one variable \( \phi  \) whereas the entropy maximization
involves the two variables \( p \) (or \( q \)) and \( \psi  \).
Moreover, the energy constraint has been absorbed. It is thus simpler
than the maximization of the entropy with two constraints. Moreover,
as we shall see in section \ref{sec:Equilibrium_Small_Rossby}, the
peculiar shape of the function \( f_{C} \), with two minima (see
figure \ref{fig_U}) will allow us to have a direct hint on the structure
of the solution. }
\selectlanguage{english}

Let us compute the equation verified by the critical states (the Euler-Lagrange
equations) of the modified free energy (\ref{Modica_Variationnel_EnergieLibre_Final}).
For this we consider small variations \( \delta \phi  \) of the functional
\( \mathcal{F}+R\alpha _{1}\mathcal{A} \), where \( -R\alpha _{1} \)
is the Lagrange parameter associated to the conservation of the area
\( \mathcal{A} \). After straightforward computations, we obtain:
\begin{equation}
\label{gibbs}
-R^{2}\Delta \phi +\phi -Rh(y)=\tanh \left( C\phi \right) -R\alpha _{1}
\end{equation}
 Let us suppose that \( \phi  \) minimize this variational problem.
Let us then define the stream function \( \psi  \) by: \begin{equation}
\label{Lien_phi_psi}
\psi =R^{2}\left( \phi +\alpha \right) 
\end{equation}
If we use the relation (\ref{Lien_phi_psi}), setting \( \alpha -R\alpha _{1}=B \),
we obtain the following equation: \( q=-\Delta \psi +\psi /R^{2}-Rh(y)=B+\tanh \left( C\left( \psi /R^{2}-\alpha \right) \right)  \).
This equation is also the critical point of maximization of the entropy
(\ref{Probleme-Variationnel-Entropie}) (see Bouchet and Sommeria
2002), where \( C\alpha  \) and \( \beta =-C/R^{2} \) are the Lagrange
parameters associated to the conservation of the area and of the energy
respectively. This equation describes a stationary solution of the
Quasi-Geostrophic equation. These two equations, for the stream function
\( \psi  \), or for \( \phi  \) (\ref{gibbs}) will be called the
Gibbs state equations. \\
We have shown that the critical points of the modified free energy
(\ref{Modica_Variationnel_EnergieLibre_Final}) are also critical
point of the maximization of the entropy under constraint (\ref{Probleme-Variationnel-Entropie}).
However, this does not prove that the \emph{minima} of the free energy
are actually \emph{maxima} of the entropy under the constraints. It
can be proven on a very general ground (see Bouchet and Barré 2003)
that any minimum of the free energy \( F=S/\beta -E \), with the
area constraint, is a minima of the entropy with energy and area constraint
(\ref{Probleme-Variationnel-Entropie}) (the converse in wrong in
general). We can thus study the minimization of the free energy, and
verify afterwards that all the possible energy are obtained. This
will be the case in our study, in the limit of small Rossby deformation
radius. To prove that the minimization, with area constraint, of the
free energy and of the modified free energy (\ref{Modica_Variationnel_EnergieLibre_Final})
are equivalent, one can either explicitly study the stability of the
solutions or prove that these two variational problem are equivalent
to a third one with two independent variables \( \psi  \) and \( \phi  \).
This point is addressed in in Bouchet (2001), proving that minima
of the modified free energy are maxima of the entropy with constraints.

\subsection{The limit of small Rossby deformation radius.}

\label{sec:Equilibrium_Small_Rossby}

In this section, we analyze the solution of the minimization of the
modified free energy \( \mathcal{F} \) (\ref{Modica_Variationnel_EnergieLibre_Final})
with the constraint of the area \( \mathcal{A} \) , in the limit
of small deformation radius. We will always consider \( C>1 \). 

We have to minimize the functional \( \mathcal{F}=\int _{D}d{\bf r}\, \left[ \frac{R^{2}\left( \nabla \phi \right) ^{2}}{2}+f_{C}(\phi )-R\phi h(y)\right]  \)
with a constraint. The modified free energy reduces, at lower oder
in \( R \), to the minimization of \( \int _{D}d{\bf r}\, f_{C}(\phi ) \).
The actual shape of the function \( f_{C} \) will therefore be essential
to the discussion. Figure \ref{fig_U} shows this shape when \( C>1 \)
(see \ref{Modica_Variationnel_EnergieLibre_Final} for the definition
of \( f_{C} \)). This figure shows that \( f_{C} \) is even and
possess two minima that we shall denote \( \pm u \). \( u \) verify:
\begin{equation}
\label{uu}
u=\tanh (Cu).
\end{equation}
The minimization of this functional, without the topography term,
also represents the coexistence of two phases in a situation of first
order phase transitions in classical thermodynamics (Van-Der-Walls
Cahn-Hilliard model). Let us discuss it, for instance, for a coexistence
between a gas and a liquid phases. The two minimum value of the volume
free energy \( f_{C} \) then correspond to the specific volume of
each phase at equilibrium. The constraint then fixes the respective
volume occupied by the two phases. When first order terms are considered,
the gradient term describes the transition surface between one phase
to the other. A surface free energy is then associated to this transition.
The minimization of this surface free energy then leads to bubble
for equilibrium structures. 

We note that the mathematical study of functional of the type \ref{Modica_Variationnel_EnergieLibre_Final},
but without the topography, is considered in Modica (1987). The functional
analysis study of this work, prove the hypothesis at the base of this
qualitative description: \( \phi  \) will take the two values \( \pm u \)
in subdomains separated by transition area of width scaling with \( R \).
Using this, we will propose a very intuitive asymptotic expansion
to describe the solutions (please note that our problem is two-dimensional,
surfaces will be replaced by lines). With respect to Modica work,
our expansion will allows a precise description of the jet, and the
generalization of the results when a topography term is taken into
account. A more complete and satisfying description of the whole asymptotic
expansion, generalizable at all order in \( R \), with mathematical
justification of the existence of the solutions for the jet equation
at all order in \( R \), is provided in Bouchet (2001). Higher order
effects are also discussed in this work. We now present a simplified
discussion.

For vortices, the two phases correspond to two different value of
the mixing of the PV. The conservation of the volume corresponds to
the conservation of the global PV. As we will see, the effect of the
topography will lead to a balance between the minimization of the
length free energy and of the tendency of positive PV to stand around
extrema of the topography, leading to the very characteristic elongated
shape of Jovian vortices.

\subsubsection{The zeroth order stream function outside of the jet: a quiescent
core }

\begin{figure}

\caption{At zeroth order, \protect\( \phi \protect \) takes the two values
\protect\( \pm u\protect \) on two subdomains \protect\( A_{\pm }\protect \).
These subdomains are separated by strong jets. The actual shape of
the structure, or equivalently the position of the jets, is given
by the first order analysis.}

\label{fig_domaine}

\par\centering \resizebox*{12cm}{!}{\includegraphics{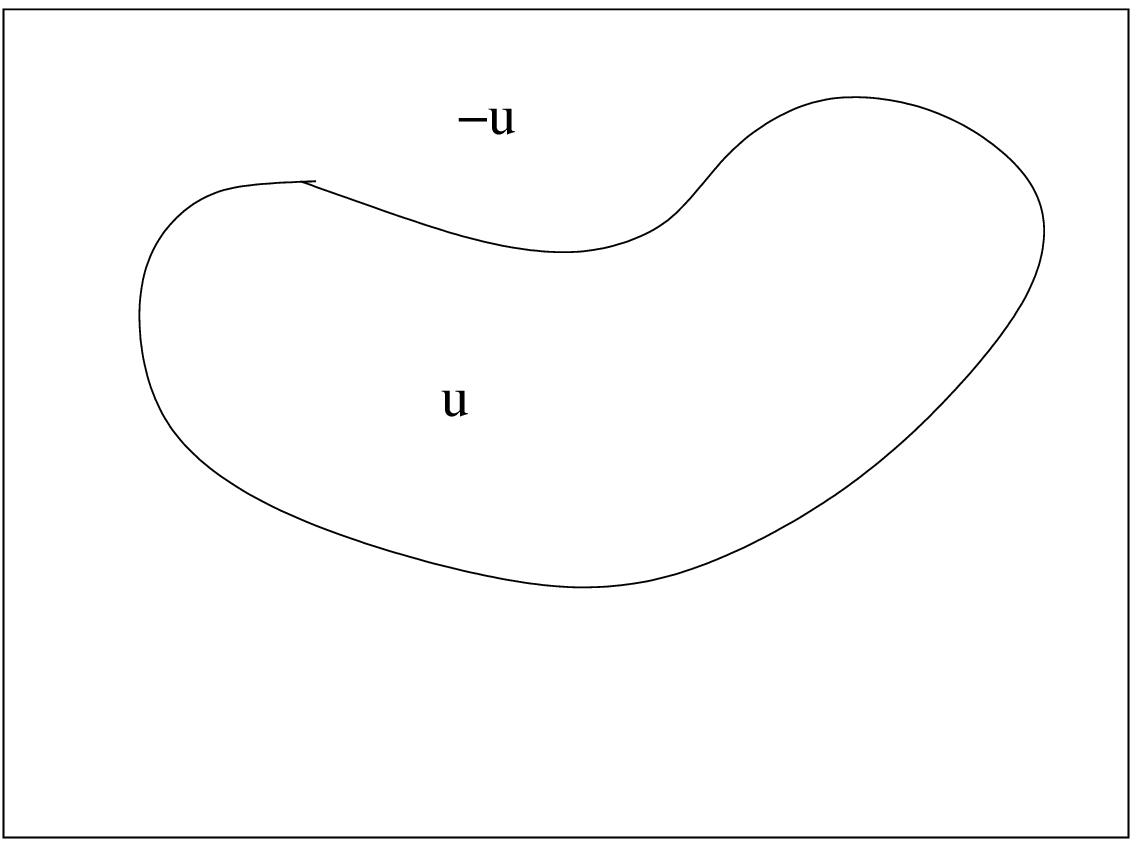} 
} \par{}
\end{figure}

At lower order the value of \( \phi  \) will therefore take the two
values \( u \) and \( -u \) in two subdomains, denoted respectively
\( A_{+} \) and \( A_{-} \), as illustrated in figure (\ref{fig_domaine}).
The constraint \( \mathcal{A}\left[ \phi \right] =\int _{D}d{\bf r}\, \phi =-\alpha  \),
taking \( A_{+}+A_{-}=1 \), will determine the respective area occupied
by these two values: \( 2A_{\pm }=\left( 1\mp \alpha /u\right)  \).
This implies \( u>\alpha  \). The actual subdomain shape will be
determined by the second order analysis. At this stage the two domains
\( A_{+} \) or \( A_{-} \) may also not be connected.\\

The above discussion solve the first order problem. Using the link
between \( \phi  \) and the stream function (\ref{Lien_phi_psi}),
we can compute the first order values of \( \psi  \): \( \psi _{\pm }=R^{2}\left( \pm u+\alpha \right)  \),
the corresponding value of \( B \): \( B=-\int _{D}d{\bf r}\, \left( \tanh \left( C\phi \right) \right) =uA_{-}-uA_{+}=\alpha  \),
and the corresponding value of the energy (\ref{Econs}) : \( 2R^{2}E=\psi ^{2}_{+}A_{+}+\psi ^{2}_{-}A_{-} \).
This yields \( 2E=R^{2}\left( u^{2}-\alpha ^{2}\right)  \), where
\( u \) is a function of \( C \) given by (\ref{uu}). \\

For sake of simplicity, we parameterize the state by the two parameters
\( u \) and \( B \), with \( u>B \). We thus obtain, at lower order
in \( R \) :

\begin{equation}
\label{psiu}
\psi _{\pm }=R^{2}(B\pm u)
\end{equation}
 \begin{equation}
\label{aire}
A_{\pm }=\frac{1}{2}\left( 1\mp \frac{B}{u}\right) 
\end{equation}
 and \begin{equation}
\label{eneu}
E\simeq E_{A}=\frac{R^{2}}{2}(u^{2}-B^{2})+\mathcal{O}\left( R\right) 
\end{equation}
 Therefore all the quantities are determined from the asymmetry parameter
\( B \) and from the parameter \( u \), related to the energy by
(\ref{eneu}). \\

In the limit of low energy, \( u\rightarrow |B| \), when for instance
\( B>0 \), then \( A_{1} \) goes to zero, so that \( \psi _{-1} \)
tends to occupy the whole domain. This state is the most mixed one
compatible with the constraint of a given value of \( B \) (or equivalently
a given initial patch area \( A=(1-B)/2 \)). In the opposite limit
\( u\rightarrow 1 \), we see from (\ref{psiu}) that in the two subdomains
\( q=\psi /R^{2} \) tends to the two initial PV levels \( a_{1}=1+B \)
and \( a_{-1}=-1+B \). Thus, this state is an unmixed state. It achieves
the maximum possible energy \( E=\frac{R^{2}}{2}(1-B^{2}) \) under
the constraint of a given patch area. We conclude that the parameter
\( u \), or the related 'temperature' \( C_{0} \), characterizes
the mixing of these two PV levels. We shall call \( u \) the segregation
parameter, as it quantifies the segregation of the PV level \( a_{1} \)
(or its complementary \( a_{-1} \)) between the two phases.

\subsubsection{The strong jet equation}

As stated before, the preceding analysis does not take into account
the interface between the two area \( A_{\pm } \). At this interface,
the value of \( \phi  \), and thus of the stream function will change
rapidly, on a scale of order \( R \). It will thus corresponds to
a strong jet. To analyze this interface, we consider the Gibbs states
equation (\ref{gibbs}) at the lower order in \( R \). Using that
the interface develops on a length scale of order \( R \), we consider
\( \tau  \) the coordinate normal to the jet, rescaled by \( R \):
\( \zeta =R\tau  \) . In the Laplacian term, we neglect the curvature
which has to be taken into account at the first order in \( R \).
We then obtain:\begin{equation}
\label{jet}
-\frac{d^{2}\phi }{d\zeta ^{2}}=\tanh \left( C\phi \right) -\phi =-\frac{d}{d\phi }f_{C}\left( \phi \right) 
\end{equation}
 The jet equation is thus the equation of a particle position \( \phi  \)
in a potential \( -f_{C}\left( \phi \right)  \). As \( f_{C}\left( \phi \right)  \)
has exactly two minima: \( f\left( \pm u\right)  \), there is a unique
trajectory with limit conditions \( \phi \rightarrow \pm u \) for
\( \zeta \rightarrow \pm \infty  \). 

This analysis shows that the jet scales typically as the Rossby deformation
radius. Moreover, in dimensionless units, the jet width and the jet
maximum velocity are given by \( Rl\left( u\right)  \) and \( Rv_{max}\left( u\right)  \)
where \( l \) and \( v_{max} \) are functions of \( u \) that may
be numerically tabulated from the resolution of the previous differential
equation. These relations allow to compute \( u \) from the jet properties.

\subsubsection{The first order stream function outside of the jet: the weak shear
flow}

To determine the first order stream function outside of the jet, we
consider the Gibbs state equation (\ref{gibbs}) by neglecting the
Laplacian term. We thus obtain the algebraic equation: \( \phi -Rh(y)=\tanh (C\phi )-R\alpha _{1} \).
Using that \( \phi =\pm u \) at zero order, we calculate the first
order solution to this equation. Using the results (\ref{uu}), this
yields \( \phi =\pm u+R\delta \phi =\pm u+R\left( h(y)+\alpha _{1}\right) /\left( 1-C(1-u^{2})\right)  \).
Using the link between \( \phi  \) and the stream function (\ref{Lien_phi_psi})
and the link between the stream function and the velocity (\ref{u}),
we obtain:\begin{equation}
\label{shear}
{\textbf {v}}=R^{3}\left( \frac{dh/dy}{1-C(1-u^{2})}\right) {\textbf {e}_{x}}
\end{equation}
This relates the zonal flow outside of the jets, the topography, and
the parameter \( u \) (determined from the total energy or from the
jet properties).

\subsubsection{Determination of the vortex shape: the typical elongated shape\label{sec_Vortex_Shape}}

Until now the jet shape has not been determined. To determine it we
can alternatively consider the jet equation at first order or compute
the first order modified free energy \( \mathcal{F} \) (\ref{Modica_Variationnel_EnergieLibre_Final}).
We make this second choice as it will enlightened the interpretation
of the solution. 

The zero order modified free energy \( \mathcal{F}_{0} \) could be
computed from the value of \( \phi =\pm u \) outside of the jet and
from the areas \( A_{\pm } \) (\ref{aire}). As it is not of interest
for the following discussion, we don't do it explicitly. Let us call
\( R\delta \phi  \) the first order modification of \( \phi  \),
computed in the previous section. It gives the contribution \( R\int _{D}d{\bf r}\, df_{C}(\pm u)\delta \phi  \)
to the modified free energy, at first order. But as by definition
of \( u \), \( df_{C}(\pm u)=0 \), this contribution is null. Let
us compute the first order contribution of the topography term \( RH=\int _{D}d{\bf r}\, \left( -R\phi h(y)\right)  \).
For this we use the zero order result \( \phi =\pm u \). We then
obtain \( H=H_{0}-2u\int _{A_{+}}d{\bf r}\, h(y) \), where \( H_{0}\equiv u\int _{D}d{\bf r}\, h(y) \).
We note that \( H_{0} \) does not depends on the jet shape. The last
contribution comes from the jet. As the jet scales like \( R \),
to determine the jet contribution at first order in \( R \), we only
need the jet determination at zeroth order. It is then easy to convince
oneself that the jet contribution is proportional to its length \( L \):
it will have the form \( Re\left( u\right) L \) where \( e(u) \)
depends only on the zeroth order jet property, itself described by
(\ref{jet}). One can then obtain \( 2e\left( u\right) =\int _{-\infty }^{+\infty }\left( \frac{d\phi }{d\tau }\right) ^{2}d\tau >0 \),
where \( \phi \left( \tau \right)  \) is a solution of the jet equation.

We thus obtain the first order expression for the modified free energy
functional:

\begin{equation}
\label{Energy_libre_ordre1}
\mathcal{F}=\mathcal{F}_{0}+RH_{0}+R\left( e\left( u\right) L-2u\int _{A_{+}}h(y)d^{2}{\textbf {r}}\right) 
\end{equation}
We compute on the same way the first order area \( \mathcal{A} \).
As the jet solution \( \phi \left( \tau \right)  \) is even, if we
choose \( \phi \left( 0\right) =0 \), there will be no jet contribution
to \( \mathcal{A} \) at this order. The only contribution will come
from the first order modification of \( \phi  \) outside of the jet
and will therefore be independent of the jet shape. We thus conclude
that, at this order, the minimization of the free energy with the
area constraint is equivalent to the minimization of (\ref{Energy_libre_ordre1})
with a given area \( \mathrm{A}_{+} \). As the two first terms of
(\ref{Energy_libre_ordre1}) are constant, this new variational problem
is a variational problem on the shape of the jet.\\

\begin{figure}

\caption{Phase diagram of the Gibbs states versus the energy \protect\( E\protect \)
and the asymmetry parameter \protect\( B\protect \) (see (\ref{B})),
when no topography is present (\protect\( h=0)\protect \). The outer
solid line is the maximum energy achievable for a fixed \protect\( B\protect \)
: \protect\( E=\frac{R^{2}}{2}(1-B^{2})+\mathcal{O}\left( R^{3}\right) \protect \).
Straight jets are obtained for the nearly symmetric cases (\protect\( B\protect \)
around 0), while a vortex is formed when one of the PV levels has
a lower area. This vortex takes the form of a circular jet for sufficiently
high energy. The frontiers line between the straight jets and the
circular jet is determined by the minimization of the jet length (first
order free energy). The hashed line represents the energy value for
which vortex area \protect\( A_{1}\protect \) or \protect\( A_{-1}\protect \)
(\ref{aire}) is equal to \protect\( (2l)^{2}\protect \), where \protect\( l\protect \)
is the typical jets width. At the left of this line, the small Rossby
deformation radius asymptotic expansion is no more valid. For such
case, asymmetric equilibrium have been described in Bouchet and Sommeria
(2002). This hashed line depends on the value of \protect\( R\protect \),
the ratio of the Rossby deformation radius to the domain scale. It
has been here numerically calculated for R = 0.03.}

\label{phase}

{\centering \resizebox*{0.8\textwidth}{!}{\includegraphics{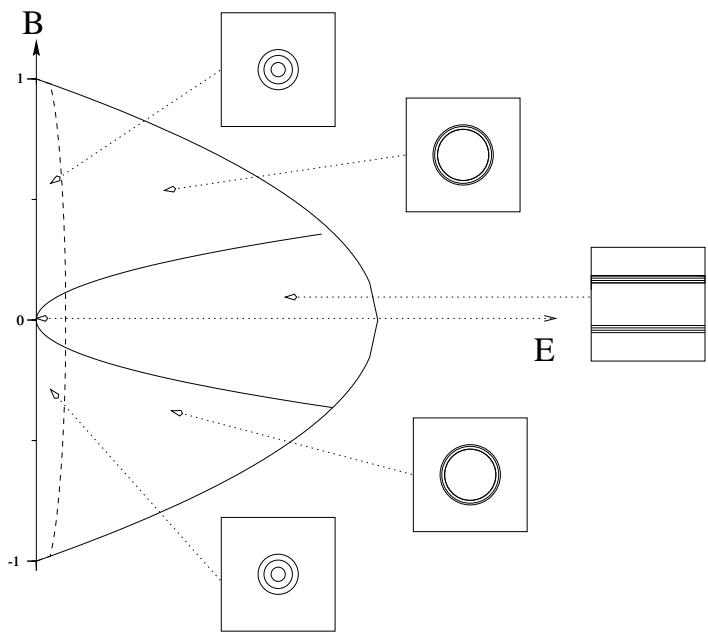}} \par}
\end{figure}

On one hand, if the topography is zero at first order, we observe
that this variational problem corresponds to the minimization of the
jet length for a given area. The jet are then straight (zonal bands
separated by strong jets) or circular (circular vortex) (this is the
equivalent of bubbles in first order phase transitions). Figure (\ref{phase})
shows the corresponding phase diagram with respect to the energy and
to the asymmetry parameter \( B \). On the other hand when a topography
is present at first order, the tendency to minimize the jet length
will be counterbalanced by the second term: the positive (resp negative)
PV tends to concentrate on extrema (resp minima) of the topography.
For a topography \( h\left( y\right)  \), the vortices will therefore
be elongated in the zonal direction. 

To give a quantitative description of this fact, we can obtain from
the variational problem (\ref{Energy_libre_ordre1}), an equation
describing the shape of the vortex. We refer to Bouchet and Sommeria
(2002) appendix B, for the actual computation. The result is that
the radius of curvature \( r \) of the curve formed by the jet (for
instance the curve of value \( \phi =0 \)) must verify:\begin{equation}
\label{Rayon_courbure}
\epsilon u\left( h(y)-\alpha _{1}\right) =e(u)\frac{1}{r}
\end{equation}
where \( \varepsilon =1 \) (resp \( \varepsilon =-1 \)) for an anticyclone
(resp cyclone) solution, and \( \alpha _{1} \) is a Lagrange parameter
associated to the conservation of the area \( \mathcal{A} \). This
relates the vortex shape to the topography and parameter \( u \).
As said in the introduction of this section it is also possible to
obtain such a result from the first order jet equation analysis. \( \alpha _{1} \)
then turns out to be the Lagrange parameter appearing in the Gibbs
state equation (\ref{gibbs}). \\

\begin{figure}

\caption{Top: typical vortex shape obtained from the curvature radius equation
(\ref{Rayon_courbure}) for two values of the parameters (arbitrary
units). This illustrate the very characteristic particularity of Jupiter's
vortices to be vary elongated, ones they reach an extremal latitude
\protect\( y_{m}\protect \) (\ref{ym}). Bottom left: the Great Red
Spot and one of the White Ovals. Bottom right: one of the Brown Barges
of Jupiter's north atmosphere. This shows that equilibrium structures
are able to reproduce the characteristic and peculiar elongation of
jovian vortices.}

\label{fig_ellipse}

\begin{tabular}{c}
\begin{tabular}{c}
\resizebox*{!}{0.5\textheight}{\includegraphics{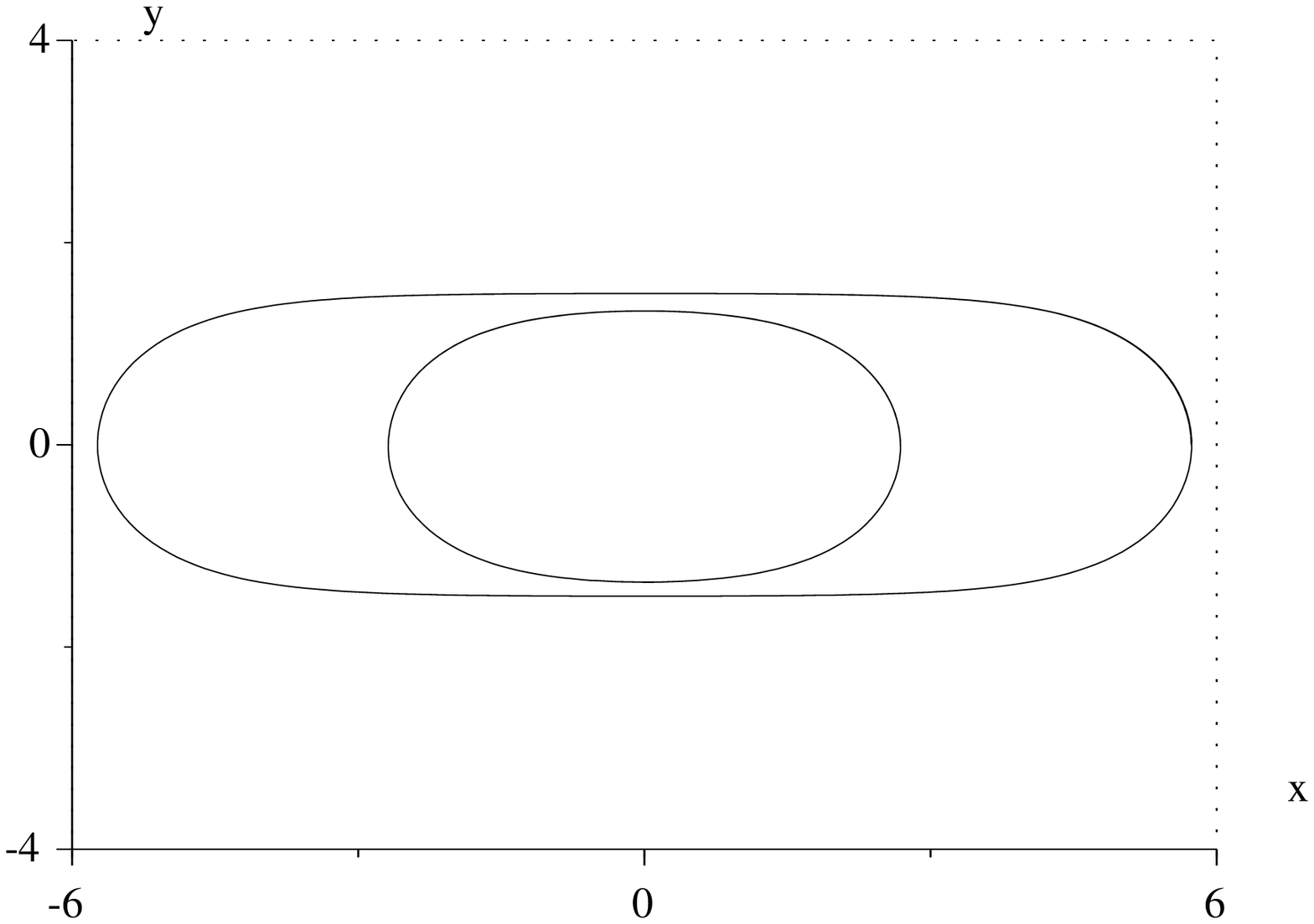}} \\
\begin{tabular}{cc}
\resizebox*{0.5\textwidth}{!}{\includegraphics{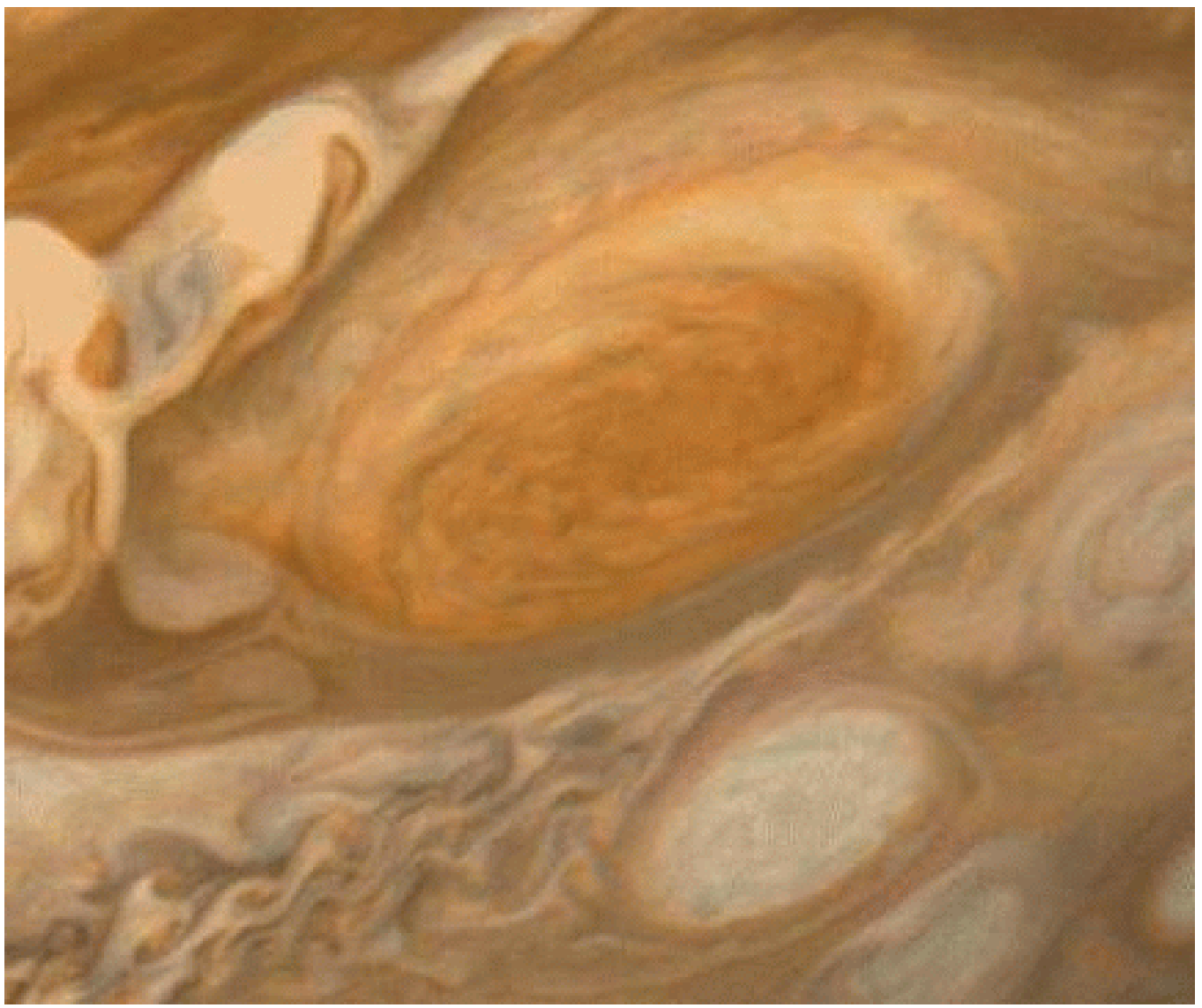}} &
\resizebox*{0.5\textwidth}{!}{\includegraphics{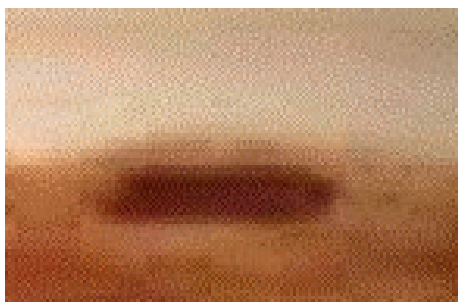}} \\
\end{tabular}\\
\end{tabular}\\
\end{tabular}
\end{figure}

In appendix \ref{appendice_forme_jet} we give equations which permit
to numerically compute the vortex shape, from equation \ref{Rayon_courbure}.
Figure \ref{fig_ellipse} compares the numerically obtained vortex
shapes, with the Jovian ones. This shows that the solution to equation
(\ref{Rayon_courbure}) have the typical elongated shape of Jovian
vortices. 

In appendix \ref{appendice_forme_jet}, we also analytically compute
the half width of the vortices \( y_{m} \). For the cosine topography
(\ref{Topographie}), we obtain :\begin{equation}
\label{ym}
y_{m}=\frac{1}{2}g\left( \frac{e\left( u\right) }{2au}\right) 
\end{equation}
 where \( g \) is the inverse of the function \( \sin x-x\cos x \)
for \( 0<x<\pi  \). 

From this formula, we see that the maximal latitudinal extension of
the vortex solution is given by a typical length defined by the topography.
We stress the important point that this maximal extension \( y_{m} \)
is independent on the parameter \( B \), or equivalently is independent
on the area of the vortex. Varying this area, the only way for the
vortex to extend is to be very elongated in the zonal direction. This
very strong qualitative property of these equilibrium solutions is
in agreement with the observed brown barges, as illustrated by figure
\ref{fig_ellipse}.

\begin{figure}

\caption{QG topography (units \protect\protect\protect\protect\protect\( s^{-1}\protect \protect \protect \protect \protect \))
versus latitude computed using the data of Dowling and Ingersoll (1989)
: a) under the GRS ; b) under the Oval BC. The analysis of the velocity
data in the Quasi-Geostrophic framework, thus clearly shows extrema
of topography under these two vortices. This is in accordance with
what we deduce from the statistical equilibrium study.}

\label{Fig_Topography_Dowling}

\par\centering \resizebox*{12cm}{!}{\includegraphics{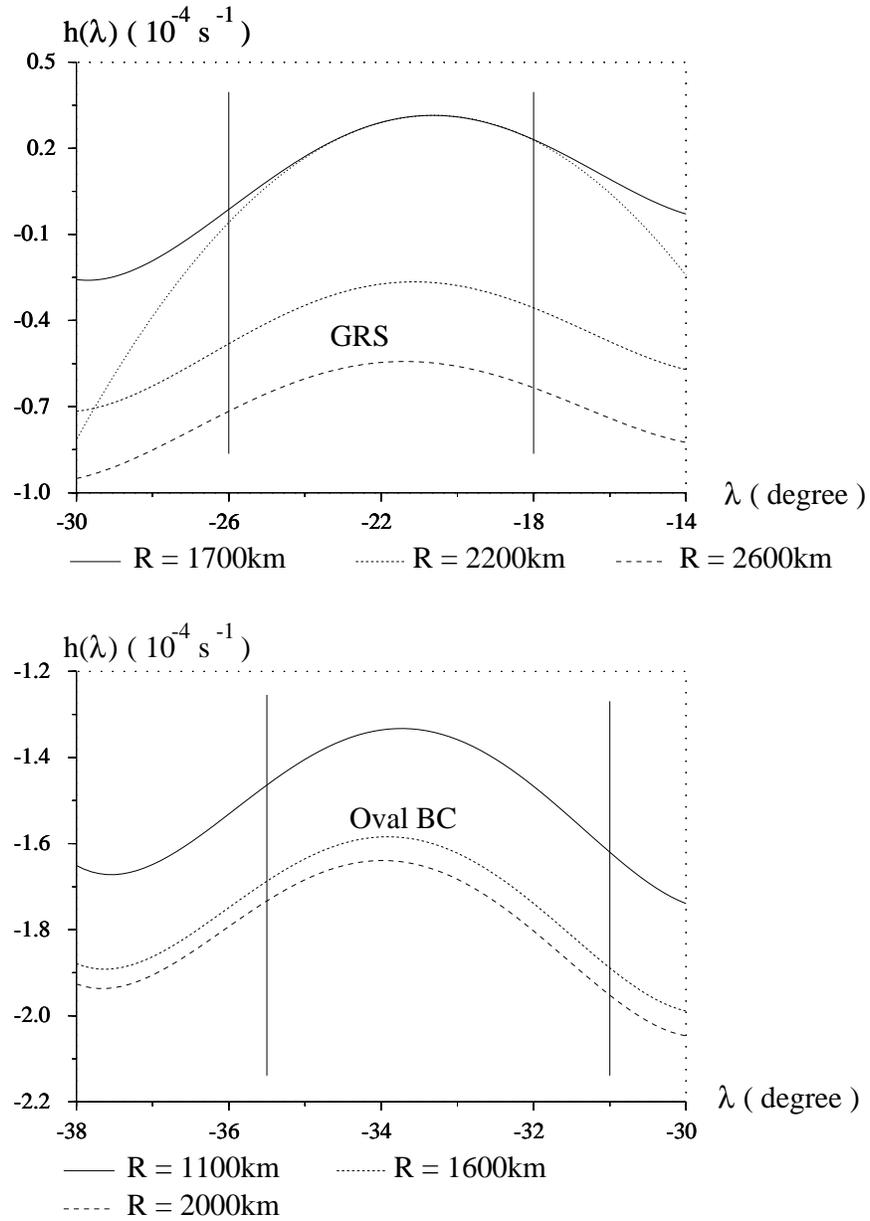} 
} \par{}
\end{figure}

A very strong property for the topography may be obtained from the
equation for the jet curvature radius. For a vortex solution, latitudinally
elongated, like the Jovian ones, the radius of curvature of the jet
have its minimum value at the latitude of the center of the spot.
From equation (\ref{Rayon_courbure}) we can then deduce that the
zonal topography must have an extrema under the center of the vortex.
Dowling and Ingersoll (1989) have analyzed the velocity field of the
GRS and of the white oval in the Shallow-Water framework. They have
then obtain the Shallow-Water topography. In Bouchet and Sommeria
(2002), we have used these result to compute the Quasi-Geostrophic
topography. Results are reported on figure \ref{Fig_Topography_Dowling}.
It clearly shows extrema of the topography under these two anticyclones.

If the effect of topography is strong enough: \( e\left( u\right) /\left( au\right) \gg 1 \),
the vortex latitudinal extension will be much smaller than the typical
variation length of the topography, we can then expand (\ref{ym})
around \( y_{m}=0 \). The maximal extension of the vortex is then
only determined by the curvature of the topography around its extremum
(quadratic approximation). Let us parameterize this curvature by \( a_{qd} \)
(\( h\left( y\right) =-\varepsilon a_{qd}\left( y-y_{0}\right) ^{2}+h\left( y_{0}\right)  \),
where \( y_{0} \) is the latitude where \( h \) reaches its extremum,
\( \varepsilon =\pm 1 \). For the cosine topography (\ref{Topographie})
\( a_{qd}=4a \). The expansion around \( y_{m}=0 \) yields:\begin{equation}
\label{ymax}
y_{max}=\left( \frac{3e(u)}{2\epsilon a_{qd}u}\right) ^{\frac{1}{3}}.
\end{equation}
\\

Using equation (\ref{Rayon_courbure}), we have described the vortex
solutions. As an alternative, zonal solutions exist, in which two
straight jets, flowing respectively eastward and westward, surround
an area of small change of \( \phi  \) (shear flow). These jet positions
may also be characterized by equation (\ref{Rayon_courbure}), when
\( 1/r=0 \). Their latitudinal position \( y_{\pm } \) is then determined
by \( h\left( y_{\pm }\right) =\alpha _{1} \). Their respective positions
are thus symmetric with respect to the zonal extrema of the topography.
The both type of solutions (jets or vortices) are then selected from
the maximum entropy criteria, or equivalently maximizing the values
obtained for the first order modified free energy (\ref{Energy_libre_ordre1}).
This selection depends on the parameters \( a \) defining the topography
(\ref{Topographie}), on the domain aspect ratio, and on the two parameters
\( u \) and \( \alpha  \) (equivalently the energy \( E \) and
the asymmetry parameter \( B \)). In figure (\ref{phase_top}), we
show the obtained phase diagram for a given value of \( a \) and
aspect ratio, in the quadratic topography approximation. It shows
a transition from vortex to jet solutions, when the asymmetry parameter
is sufficiently close to zero. The aspect ratio of the vortices is
also represented. We refer to Bouchet and Sommeria 2002 for a detailed
discussion. \\

In the limit of small Rossby deformation radius, the maxima of entropy
for a given PV distribution and energy, are formed by strong jets,
limiting areas characterized by a weaker shear. Straight jets forming
bands and zones, or vortex solutions, are both possible. The maximum
entropy principle allows to select the type of solution for given
parameters. The deep layer shear, the active layer shear, the shape
of the vortices, and the strength of the jets are linked by the relations
(\ref{shear},\ref{ymax} and \ref{jet}).

In next section (\ref{sec_Application_Jupiter}) we discuss application
of these results to the Jovian troposphere.

\subsection{Jovian troposphere applications\label{sec_Application_Jupiter}}

In the previous sections, we have deduced all the flow properties
for the statistical equilibrium, in the limit of small deformation
radius. The qualitative properties of the vortex solutions are the
one of the Great Red Spot : an annular strong jet, forming an oval
shaped boundary, surrounding a quiescent core and admits a zonal shear.
To our knowledge, this is the first model having these qualitative
properties. 

In this section, we want to describe the main hypothesis of such a
model for Jovian vortices, and its limitations. We also stress the
main physical consequences of our analysis. In section \ref{sec_Modele_Analytique_GRS}
we discuss the possibility to apply the model of the previous section
to quantitatively describe the Great Red Spot. This discussion comes
back on the hypothesis concerning the Potential Vorticity initial
distribution, and on the limitations of the Quasi-Geostrophic model.

In the previous analytical analysis, we have studied equilibrium structures
in the limit of small Rossby deformation radius. We have seen that
this hypothesis leads to strong concentrates jets. For instance, we
conclude that the actual small value of \( R \), with respect to
the width of the Great Red Spot is responsible for its annular structure.
For the case of other vortices, such as the White Ovals, or the north
hemisphere Brown Barges, the small radius of deformation this limit
is no more valid. In section \ref{sec:Emergence_Num_OBC}, \ref{sec:Emergence_Num_GRS}
and \ref{sec:Emergence_Num_Jets} numerical computations of the maxima
of entropy under constraints, which reproduce the properties of these
vortices. Nevertheless, the analytical analysis permits to qualitatively
understand the dependence on the parameters of such vortices, and
it has permit us to determine the parameter values for such vortices.

We discuss in section \ref{sec:zonal_topographie} some properties
the zonal upper layer shear and the deep shear, obtained in the previous
analysis, that should be of interest for Jovian vortices. We explain
why such structures are obtained only for very energetic flows. In
section \ref{sec:Rhine} we discuss an alternative to the Rhine's
scale to explain the typical vortex width. We conclude this section
by recalling the main hypothesis of the statistical mechanics approach.

\subsubsection{The effect of the Potential Vorticity distribution on the equilibrium
structures\label{sec_PV_Distribution}}

A fundamental issue is the hypothesis we made on the PV distribution.
We have supposed an initial PV made of two types of PV. As argued
in Bouchet and Sommeria (2002), this is a natural hypothesis in the
context of a Jovian latitudinal band. Indeed, this could be the result
of intense incoming thermal plumes, as recently observed by Ingersoll
et al (2000): conservation of the absolute angular momentum during
the radial expansion leads to a strong decrease of the local absolute
vorticity, which comes close to zero. This means that in the planetary
reference frame, a local vorticity patch with value \( -f_{0} \)
(the planetary vorticity) is created. The opposite vorticity is globally
created by the subducting flow, but it is close to 0 due the much
larger area. 

Of course, even if a two-level approximation is natural, the real
fine-grained PV distribution is not actually known, and an important
issue is to study the dependence of the results on such a distribution.
The knowledge of this distribution is equivalent to the knowledge
of an infinity of constraints, the Casimirs. This is a major practical
limitation of such a statistical mechanics approach (it is not a theoretical
one). A natural way to proceed is to study a-posteriori the choice
of the distribution, by comparison with observed flows. This is the
way we have proceed, by studying the simplest case, the two PV level
case, and by comparing the results to the Jupiter's structures. In
the same spirit, Turkington, Majda and DiBattista (2001) have proposed
to study a fine-grained PV distribution the centered gamma distribution,
in order to study the effect of a skewness to the PV distribution.
They have shown the importance of an anticyclonic skewness to obtain
anticyclonic structures. This is consistent with our 2 PV levels description
with \( B>0 \), and with the observed anticyclonic forcing by incoming
thermal plumes, as discussed above. Using this centered gamma distribution,
they have obtained the oval shaped vortices and jets. These jets are
not however strong jets, and they do not observe the ring structure
of the GRS. Using our study, we may explain why their distribution
is not suited to study Jovian vortices. 

The thermal plume forcing products PV patches with vorticity of order
\( -f_{0} \). Their is no physical to expect another type of forcing
to produce very large values of PV with respect to \( \left| f_{0}\right|  \).
Moreover, the fine-grained distribution is conserved, and the extrema
of the coarse-grained distribution can only decrease. This is thus
very natural to considered a PV distribution with compact support
(values of the PV bounded). This is not the case of the centered gamma
distribution, and this may have several consequences. Indeed, it can
be proved (Robert and Sommeria 1991) that for any bounded distribution,
the equilibrium relation between the PV and the stream function \( q\left( \psi \right)  \)
must be strictly monotonic and tends to two maximum \( q_{\pm m} \)
(the maximum and the minimum of the PV initial distribution) value
for \( \psi  \) going to \( \pm \infty  \). As a consequence \( q\left( \psi \right)  \)
must have at least one inflection point. The \( q\left( \psi \right)  \)
relation then has the shape of a \( \tanh  \) at infinity, possibly
with more than one inflexion point. This last property is not verified
for the centered gamma distribution. This is however an essential
property, as it is necessary condition to prove the existence of the
equilibrium structures for any parameters. Moreover, as shown by the
present study, the \( \tanh  \) like shape of \( q\left( \psi \right)  \)
(it is equivalent to the existence of at least two minima of the area
free-energy \ref{Modica_Variationnel_EnergieLibre_Final}) is essential
to obtain the phase coexistence and strong jet property of the Jovian
vortices. Moreover, besides these physical and theoretical arguments,
it corresponds to the observed \( q\left( \psi \right)  \) relation
for the GRS, as shown by the figure 12 of Bouchet and Sommeria (2002).
We thus conclude that PV distributions with compact support should
be preferentially studied to model geophysical flows. \\

The problem of the knowledge of the PV distribution is not, however,
a real limitation in the case of small Rossby deformation radius.
We will indeed argue that the main property of the equilibrium structures
are independent of the exact distribution, in this case. Let us suppose
that the initial Potential Vorticity distribution is made of an infinite
number of PV levels (not only two as supposed in the previous section),
but with bounded PV. The statistical equilibrium will then be described
by a monotonic function \( q(\psi ) \) reaching asymptotic extrema
at the minimum and maximum PV levels (Robert and Sommeria, 1991).
In most cases such a function will still be represented by a tanh
like curve (one inflection point). We still can use the minimization
of a free energy similar to (\ref{Modica_Variationnel_EnergieLibre_Final}).
The function \( f_{C} \) determining this free energy will correspond
again to the coexistence of two phases as represented in figure \ref{fig_U}.
The derivation described in the previous section is independent of
the actual shape of the function \( f_{C} \). We will then obtain
similar equations for the strong jet (\ref{jet}), surrounding shear
(\ref{shear}), curvature radius (\ref{Rayon_courbure}) and extremal
extension (\ref{ym}). Only the \( u \) depending functions in these
equations, will depend on the actual PV distribution. We recall that,
as illustrated by the figure 12 of Bouchet and Sommeria (2002), a
tanh-like shape is observed for the GRS. 

We may also imagine a curve \( q(\psi ) \) with more than one inflexion
point, instead of a single one, resulting in the coexistence of more
than two phases. The most common case will be however a two-phase
equilibrium. Likewise in usual thermodynamics the coexistence of more
than two solutal phases is unlikely, even when many chemicals (equivalent
to PV levels) are mixed. Nevertheless,  we still can use the minimization
of a free energy similar to (\ref{Modica_Variationnel_EnergieLibre_Final}).
The function \( f_{C} \) determining this free energy will then correspond
to the coexistence of three (or more) phases. Solutions can then be
an anticyclone on a topography bump coexisting with a cyclone on a
topography minima, both surrounded by a mean PV area. The jet structure
of each of these vortices will then always be described by equations
similar to the jet equation (\ref{jet}), curvature radius equation
(\ref{Rayon_courbure}) and extremal extension equation (\ref{ym}). 

We thus conclude that the qualitative structure of the statistical
equilibrium is independent of the actual PV distribution. This result
is valid only when the Rossby deformation radius is small.

\subsubsection{A quantitative model for the Great Red Spot\label{sec_Modele_Analytique_GRS}}

The relations obtained in the previous section between the topography,
the maximum jet velocity, the jet width, the surrounding shear, and
the vortex shape have been written in dimensional form, in Bouchet
and Sommeria (2002). We have then shown that the actual observed values
of these physical properties, for the Great Red Spot, can be matched
with this model. This then allowed the determination of the Rossby
deformation radius and of the topography curvature under the spot.
This proves that a model of the GRS, by a statistical equilibria of
the Quasi-Geostrophic model, with a quadratic topography, with an
initial condition made of two values of the initial potential vorticity,
can fit observations with precision. 

We discuss further these hypothesis. The first one concerns the topography.
We have shown that for any topography, the essential point is that
it has an extremum under the spot. This is confirmed by observations
(see figure \ref{Fig_Topography_Dowling}). The hypothesis of a quadratic
topography is then natural. In this article, we have studied the effect
of a cosine topography. The main results are the same. However the
actual value of the width of the vortex may be changed. Actually,
we will numerically compute the velocity field of the GRS for a cosine
topography in section \ref{sec:Emergence_Num_GRS}, and the actual
values of the Rossby deformation radius and of the topography curvature
will slightly change.

Concerning the potential vorticity distribution, we have argued in
section \ref{sec_PV_Distribution} that the qualitative description
does not depends on the actual initial PV distribution. However, the
quantitative description, for instance of the shape of the vortex,
depends on it (via the \( u \) depending functions). For instance,
we think that a model with another initial PV distribution, may also
allow to fit observations with precision, leading possibly to slightly
different values for the Rossby deformation radius or for the actual
topography curvature.\\

The validity of the Quasi-Geostrophic model, and of the description
of the Jovian troposphere by a single layer, are limitations of our
model. The validity of the Quasi-Geostrophic model, for the GRS description,
has been discussed by Dowling and Ingersoll (1989) and it was found
reasonably good as a first approach. It is not fully accurate, for
instance, the maximum value of the Rossby number has been evaluated
to be 0.36 (near the jet maximum curvature) (Mitchell and collaborators,
1981). We note that an analysis of equilibrium states in the Shallow-Water
model leads essentially to the same structure (Bouchet, Chavanis and
Sommeria, 2003), as the one presented in the present work, with corrections
due to ageostrophy.

\subsubsection{Energy, zonal shear and topography\label{sec:zonal_topographie}}

We have proven in section \ref{sec_Vortex_Shape}, that vortices are
located on topography extrema. This has been verified using the GRS
and White Oval data (see figure \ref{Fig_Topography_Dowling}). In
section \ref{sec_PV_Distribution}, we have argued that, for any PV
distribution, the relation linking the radius of curvature of the
jet with the topography will be again (\ref{Rayon_courbure}), where
only the \( u \) depending terms will be changed. Our conclusion
on the topography extrema is thus independent on the actual PV distribution.

In the following, we stress some important consequences of our analytical
analysis, concerning the shear flow, which are also independent on
the actual PV distribution. Equation (\ref{shear}) describes the
shear outside of the jets. Using (\ref{uu}), it can easily proven
that \( 1-C(1-u^{2})<1 \). Thus the shear in the active layer \( \sigma =d{\textbf {v}}_{x}/dy \)
is larger than the shear in the deep layer : \( \sigma _{d}=R^{3}d^{2}h/dy^{2} \).
Qualitatively, this may be seen as a consequence of the fact that
positive Potential Vorticity will sit predominantly on topography
bumps. 

Let us give a justification on a more general ground, in order to
argue that this result is independent on the PV distribution. . We
first prove that any \emph{statistical equilibrium, not zonal (a vortex
for instance), must have an energy \( E>0 \).} Let us consider a
statistical equilibrium for any PV distribution. On one hand, it can
be proven on a general ground, that for positive temperatures states
\( \beta >0 \), only one solution to the equilibrium state equation
exist (see for instance Michel et Robert 1994a). On the other hand,
in a periodic geometry, with topography, or in a channel geometry,
it can be proven easily that a zonal solution exists (following Michel
and Robert proof of the existence of the equilibrium, but restricting
the study to a one dimensional equation). This proves that positive
temperature states are zonal. Moreover, when only one state is possible
in such maximization of entropy with a given energy, it can be proven
that the inverse temperature \( \beta  \) is a decreasing function
of the Energy \( E \) (or equivalently that the equilibrium entropy
is concave, see Bouchet and Barré (2003) for a justification). From
this we deduce that all states with \( \beta >0 \) have an energy
lower than the state with \( \beta =0 \). The only state with \( \beta =0 \)
is a completely mixed state: \( q=0 \), thus \( \psi \left( y\right) =R^{3}h\left( y\right)  \),
and \( E=0 \). We thus conclude that all equilibrium structures with
energy \( E<0 \) are zonal. Conversely, this proves that any statistical
equilibrium, not zonal (a vortex for instance), must have an energy
\( E>0 \). We recall that we have supposed \( C=-R^{2}\beta >1 \),
in the analysis of equilibrium states (section \ref{sec:Equilibrium_Small_Rossby}).
We note that this reasoning may be easily applied to any stable stationary
state of the Quasi-geostrophic equation, with topography, in a channel
or doubly-periodic geometry (by considering the functional which is
minimized in the derivation of the first Arnold stability theorem
(Arnold 1961)). We also note that this result is independent off any
hypothesis on the PV distribution and on the value of \( R \). When
the PV distribution is known a-priori, the critical value \( \beta _{c} \)
of \( \beta  \) (resp the Energy) below (resp above) which non zonal
solutions may exist, may be proven to be strictly positive. For instance,
for the two-level distribution we have considered, it can be proven
that \( \beta _{C}>1/R^{2}+\lambda _{1} \), where \( \lambda _{1} \)is
the first eigenvalue of the Laplacian, for the geometry considered. %
\footnote{This criterion is linked with the hypothesis of the second Arnold
stability theorem.
}. 

We end this discussion, by qualitatively linking this result, with
the strength of the shear. We first note that in the Quasi-Geostrophic
model, PV interacts mainly with PV values at a distant do lower than
a typical length of order \( R \). This allows to conclude that negative
PV patches on topography trough lower the Energy (and conversely for
positive PV patches). As a consequence, in any state with \( E>0 \),
positive PV must dominate negative PV on topography bumps. The shear
in the upper layer is thus larger than in the lower layer.\\

We have proven that \emph{}stable stationary flows, not zonal (a vortex
for instance), must have an energy \( E>0 \). This has a strong practical
implication : to numerically obtain vortices like the Jovian ones,
with small values of \( R \), one has to start with an initial conditions
where positive PV dominate negative PV on topography bumps.

\subsubsection{An alternative to the Rhines' scale \label{sec:Rhine}}

For a geostrophic turbulence with a linear \( \beta  \) effect (\( h\left( y\right) =\beta y \)),
it has been argued that typical length for the vortex size should
be \( L_{\beta }=\pi \sqrt{U/\beta } \) (the Rhines' scale, Rhines
and Young 1982), where \( U \) is a typical flow velocity. On the
contrary, when the value of the Rossby deformation radius is small,
for a strictly linear beta-effect, the statistical equilibrium vortex
solution are circular, with jet width scaling with, but without limit
to their size, due to the beta-effect. The beta-effect is only responsible
for a constant westward velocity drift such as to compensate the beta-effect
(see Bouchet and Sommeria 2002).

When a more complex topography is taken into account (not linear),
our study in section \ref{sec:Equilibrium_Small_Rossby} has shown
that the vortex width has a maximal value. The maximal latitudinal
extension for a zonal topography is for instance given by (\ref{ym})
for a cosine topography or by (\ref{ymax}) for a quadratic topography.
We thus deduce from this analysis that a typical vortex width is related
to the topography curvature, and not to the topography first-derivative,
as in the Rhines' scale case. The topography curvature is itself directly
related to the shear surrounding the vortex (\ref{shear}) or to the
deep shear. We obtain the following dimensional typical latitudinal
extension for the vortex \( L_{\sigma }=\left( R^{2}U/\sigma _{d}\right) ^{1/3} \),
where \( U \) is the typical strong jet velocity, and \( \sigma _{d} \)
is the deep shear, of the same order as the shear surrounding the
vortex or. If we moreover consider that the typical potential vorticity
is of order \( \left| f_{0}\right|  \), the planetary vorticity,
and that the jet width scale with \( R \), we obtain \( U\propto R\left| f_{0}\right|  \).
This gives an other expression of the typical latitudinal extension
in terms of the Rossby deformation radius and on the shear : \( L_{\sigma }=R\left( \left| f_{0}\right| /\sigma _{d}\right) ^{1/3} \).
As \( R \), \( \left| f_{0}\right|  \) and \( \sigma _{d} \) are
independent on the initial conditions (PV distribution and energy),
the typical latitudinal extension is independent on the forcing. We
recall that the exact value of the latitudinal extension, for a given
PV distribution and energy, for a given topography, may be computed
from the small \( R \) expansion (\ref{ym} or\ref{ymax}), or numerically
for larger values of \( R \).

\subsection{Relaxation equations : a small scale turbulence parameterization\label{sec:Equations_de_relaxation}}

As discussed in the beginning of this section, the equilibrium statistical
mechanics describes the states of optimum Potential Vorticity mixing,
for a given energy and PV distribution. The dynamics of the Quasi-Geostrophic
equations (\ref{QG}) should be responsible for such a mixing. From
a numerical point of view, the correct parameterization of this mixing
is a crucial issue for the modeling of geophysical flows. Accordingly
to the ideas of statistical mechanics, Robert et Sommeria (1992) have
proposed a parameterization of turbulence, for two-dimensional or
Quasi-Geostrophic flows, based on a Maximum Entropy Production Principle
(MEPP). The corresponding equations have the property to maximize
the entropy production while conserving all the dynamical invariants.
As they converge, for infinite time, towards entropy maxima, they
have been called relaxation equations. Therefore, they can also be
used to numerically compute maxima of the entropy for given PV distribution
and energy.\\

Let us present these equations in the context of the Quasi-Geostrophic
dynamics. In the following sections we will use them for dynamical
simulations. We will show their interest, compared with other parameterizations,
to perform flow simulation (section \ref{sec:Comparaison_Navier_Stokes}).
We also use them to compute equilibrium structures that we will compare
to the actual vortices of Jupiter troposphere (sections \ref{sec:Emergence_Num_OBC},
\ref{sec:Emergence_Num_GRS} and \ref{sec:Emergence_Num_Jets}). 

Relaxation equation may consider any potential vorticity distribution
(Robert and Sommeria 1992, Robert et Rosier 1996). However, for sake
of simplicity, we consider a situation for which the initial condition
is composed only of PV patches of vorticity \( a_{1} \) and \( a_{-1} \).
This choice is in accordance with the equilibrium structure analyses,
presented in sections \ref{sec_Gibbs_states} and \ref{sec:Equilibrium_Small_Rossby}
We have argue in section \ref{sec_PV_Distribution} that the qualitative
properties of the equilibrium structures are not affected by the PV
distribution, for sufficiently small Rossby deformation radius. Once
this simplification is assumed, the relaxation equations are (Robert
and Sommeria 1992): \begin{equation}
\label{Emergence_Num_relaxation}
\frac{\partial \omega }{\partial t}+{\bf u}.\nabla \omega =\nabla .\left( \nu \left[ \nabla \omega +\beta \left( a_{-1}-\omega \right) \left( \omega -a_{1}\right) \nabla \psi \right] \right) 
\end{equation}
with \begin{equation}
\label{Emergence_Num_beta}
\beta =-\frac{\int _{D}d{\bf r}\, \nu \nabla \omega .\nabla \psi }{\int _{D}d{\bf r}\, \nu \left( a_{-1}-\omega \right) \left( \omega -a_{1}\right) \left( \nabla \psi \right) ^{2}}
\end{equation}
where \( \beta  \) is the Lagrange parameter associated to energy
conservation and \( \nu  \) is a turbulent viscosity. The first term
of the right hand side of equation (\ref{Emergence_Num_relaxation})
is a usual diffusion. The second term on the right hand side of (\ref{Emergence_Num_relaxation})
is a drift term which acts to maintain a constant energy. Accordingly
to the MEPP hypothesis, it is derived such that the entropy production
is optimal. 

In section (\ref{sec:Comparaison_Navier_Stokes}), we will consider
numerical simulation using only a viscosity, this is the usual eddy
diffusivity hypothesis. We will then show that such a parameterization
is unable to reproduce even the qualitative properties of the flow,
for very long time simulations. In both cases, relaxation equations
and eddy viscosity, we will use the minimal value of \( \nu  \),
compatible with a given resolution. We note that there is no theoretical
ground to assert that the coarse-grained dynamics should be such to
maximize the entropy production. The relaxation equations (\ref{Emergence_Num_relaxation})
are however likely to better describe the dynamics because, on one
hand, they respect the conservation laws of the inertial dynamics,
and in the other hand, they take into account the tendency towards
mixing of the system.

We note that a numerical algorithm to compute maxima of entropy under
constraints, which does not use relaxation equations, is described
in Turkington \textbf{}and Whitaker (1996).

\section{White Ovals formation from randomly distributed vortices.}

\label{sec:Emergence_Num_OBC}

In this section and in the following ones, we use the relaxation equations,
presented in section \ref{sec:Equations_de_relaxation}, to simulate
an inertial dynamics and/or to compute the statistical equilibrium
of the Quasi-Geostrophic model. 

As a first experiment, we show in section \ref{sec:OBC}, how potential
vorticity patches with random positions, lead to the formation of
several vortices, which progressively merge until forming a unique
structure. Due to the presence of a topography, these structures have
an elongated shape. The parameters have been chosen to make an analogy
with the Jupiter's White Oval flow.

In section \ref{sec:Comparaison_Navier_Stokes} we compare such a
simulation with a Direct Numerical Simulation (usual viscosity).

\subsection{Anticyclones formation from randomly distributed vortices\label{sec:OBC}}

Let us consider the evolution of an initial condition formed by anticyclonic
potential vorticity patches, randomly distributed (figure \ref{fig:Emergence_Numerique_OBC1}).
The resolution of this computation is 512x128. Parameters are \( R=0.2 \),
\( a=0.4 \), \( a_{1}=4.2 \), \( a_{-1}=-1 \). We use a diffusivity
\( \nu =1.5\, 10^{-4} \). The time step is \( \Delta t=6.13\, 10^{-3} \).

Figures \ref{fig:Emergence_Numerique_OBC1} and \ref{fig:Emergence_Numerique_OBC2}
illustrate the evolution of this initial condition, modeled by relaxation
equations. They show the coalescence of the vortices, progressively
forming several anticyclones, in a latitudinal band limited by topography.
Any of these anticyclones is then centered on the topography maxima,
located in the center of the picture. These anticyclones form local
statistical equilibrium, as illustrated the scatter-plots of the potential
vorticity versus the stream function, on figure \ref{fig:Emergence_Numerique_OBC2}. 

Time lapses between pictures of figure \ref{fig:Emergence_Numerique_OBC1}
correspond to few turnover times (16 from first to last). The local
organization is thus very rapid. On the contrary the time lapse between
the two last pictures of figure \ref{fig:Emergence_Numerique_OBC2}
is approximately of 50 turnover times. During this time, the two anticyclone
have progressively achieved a local equilibrium as illustrated by
the sharpening of the two curves on the scatter plot of potential
vorticity versus stream function. Their respective position is however
quite unchanged. 

Let us recall that the deformation radius value is \( R=0.2 \), which
is very small compared to the latitudinal band length : \( 4\pi  \).
As the interaction between the two vortices decrease exponentially
for values greater than \( R \), it is in this case very small. This
explains the very long time needed for the system to achieve the exact
equilibrium structure. After a much greater time lapse (approximately
300 turnover times) the two anticyclones finally coalesce, to form
a unique structure, visible on figure \ref{fig:Emergence_Numerique_OBC3}.
\\

\subsection{The White Ovals evolution and structure}

The three White Ovals at \( 33^{o} \) S, called BC, DE and FA, where
the larger anticyclones on Jupiter, after the GRS. They formed when
an anticyclonic zone broke into three parts in 1939-40 (see Ingersoll
and collaborators 2002 for references). In 1998, the anticyclones
BC and DE merge into a larger one. This new oval then merge with the
oval BA in 2002. 

This behavior is predicted by the statistical mechanics. The quick
organization into oval shaped vortices, followed by a very long time
before the three ovals actually merge in a single structure is very
similar to the one described in the previous computation (figures
\ref{fig:Emergence_Numerique_OBC1}, \ref{fig:Emergence_Numerique_OBC2}
and \ref{fig:Emergence_Numerique_OBC3}). In this last case, these
vortex have emerged from random initial patches. However as illustrated
in section \ref{fig:Emergence_Numerique_Jet_Tache}, the same structures
could have been obtained from the destabilization of jets. 

The equilibrium velocity field (figure \ref{fig:Emergence_Numerique_OBC3})
then shows a structure very similar to the white ovals ones: the anticyclone
is too small for the limit of small Rossby radius \( R \) to apply.
As a consequence we do not observe a quiescent inner region as is
the case for the Great Red Spot. The anticyclone has nevertheless
an oval shape, linked to the deep flow shear and to the upper layer
shear. We have not given dimensional values for the quantitative characteristics
of this equilibrium structure, nor tried to match them by choosing
appropriate values of \( R \) and of the topography curvature. This
may however be done, using an iterative scheme, as we will describe
in the following section, for the Great Red Spot.

Such a work would be of special interest, in order to try to use the
observation data from the observation of the merger of these anticyclones
(see for instance Sanchez-Lavega and collaborators 2000). One could
first put some constraint on the physical parameters by using the
actual properties of the spot before merging, and verify the large
anticyclone after merging is compatible with statistical mechanics
predictions.

\subsection{Comparison of relaxation equations with usual eddy-diffusivity parameterization }

\label{sec:Comparaison_Navier_Stokes}

Figure \ref{fig:Emergence_Numerique_Comparaison_relaxation_NS} shows
vorticity fields obtained after 25 turnover times, either using eddy-diffusivity
or the relaxation equations, using in both cases the same resolution
512x128. Even if the time elapsed from the beginning of the computation
is very small compared to the global organization time, this figure
shows important qualitative differences between these two modelings.
The vorticity patches are far less compact for eddy-diffusivity type
computations. Moreover the decrease of energy is already important
for this last computation. These differences are crucial for long
time dynamics : the eddy-diffusivity type computation indeed rapidly
leads to a complete energy dissipation. As a consequence, a numerical
experiment, with an eddy diffusivity, showing the formation of anticyclones
from random initial vorticity patches and their vary slow evolution
towards a final unique vortex (as show on figures \ref{fig:Emergence_Numerique_OBC1}
and \ref{fig:Emergence_Numerique_OBC2}) is probably infeasible. 

As explained above, this is mainly due to the small value of the deformation
radius \( R \), for which the dynamical organization is very slow.
This illustrates very clearly the interest of the relaxation equations
in such a context.

\section{The Great Red Spot of Jupiter. }

\label{sec:Emergence_Num_GRS}

\subsection{A model of the Great Red Spot}

Let us propose a model of the Great Red Spot of Jupiter, as a statistical
equilibrium structure. We model the latitudinal band of the Great
Red Spot as a periodic domain of latitudinal extension \( L^{\star }=18\, 800\, km \)
and longitudinal extension \( 4X18\, 800\, km \), with a zonally
periodic topography of the same periodicity. As explain in section
\ref{sec:zonal_topographie}, the organization of the structure is
essentially local and determined by the topography under the vortex.
Thus the artificial boundary conditions used here, are of no importance
(due to the small value of \( R \), the equilibrium structure for
a wider and more elongated latitudinal band will be only slightly
different from the one computed here). We use the following parameters
: \( R^{\star }=1460\, km \), \( a^{\star }=1.3\, 10^{-16}km^{-3}s^{-1} \),
\( a^{\star }_{1}-a^{\star }_{-1}=2.14\, 10^{-4}s^{-1} \) (the corresponding
dimensionless parameters are \( R=0.234 \), \( a=0.117 \), \( B=0.87 \),
\( u=0.99 \)). We numerically compute the equilibrium structure corresponding
to these parameters, using the relaxation equations described in section
\ref{sec:Equations_de_relaxation}. 

Figure \textbf{}\ref{fig:Emergence_Numerique_tache_rouge} shows the
potential vorticity and the velocity field for the equilibrium structure,
as well as the velocity field obtain from Voyager data analysis (from
Dowling and Ingersoll 1998). Let us note the very good qualitative
agreement between the two velocity fields. We have numerically computed
the parameters of this vortex : the maximum jet velocity is \( v^{\star }_{max}=120\, ms^{-1} \),
the jet width (length between the two points where the jet velocity
is half of the maximum jet velocity) is \( l^{\star }_{x}=3600\, km \)
for the jet at mid latitude (flowing northward or southward) and \( l^{\star }_{y}=2900\, km \)
for the extremal latitude jet (flowing eastward or westward), the
maximum latitudinal extension (length from the center of the vortex
to the point northward, where the jet achieves its maximal speed)
is \( y^{\star }_{m}=3800\, km \), the aspect ratio of the spot (the
length is measured using maximum jet velocity point, as for \( y^{\star }_{m} \))
is \( \delta =1.8 \), and the surrounding shear is \( \sigma ^{\star }=0.5\, 10^{-5}\, s^{-1} \)
\\

All these quantities are compatible with the observed ones (data from
Mitchell and all 1981 analysis), except for the surrounding shear
whose real value is \( \sigma ^{\star }=1.5\, 10^{-5}\, s^{-1} \)
. We thus conclude that the statistical equilibrium of the 1-1/2 Quasi-Geostrophic
model, with a cosine topography and with a 2 level PV distribution,
allows to model quantitatively all the main characteristics of the
Great Red Spot, except for a factor \( 3 \) for the shear.\\

A natural question is whether this result could be improved in the
context of the 1-1/2 Quasi-Geostrophic model. To obtain the above
parameters, we have used the indications given by the relations (\ref{jet},\ref{shear},\ref{ym})
and the computation of the maximum jet velocity, in order to design
an iterative scheme to find the parameter best suited to the modeling
of the Great Red Spot. As this scheme converged, we do not think that
it could be better with the same topography and the same PV levels
distributions. An analysis of equations describing the vortex shape
(\ref{Rayon_courbure},\ref{ym}) either for a quadratic or for a
cosine topography, shows that a quadratic topography should give better
results. This is consistent with the study of this last case in Bouchet
and Sommeria (2002). Concerning the choice of the PV distribution,
we have argued in section \ref{sec_PV_Distribution} that a different
PV distribution, compatible with a \( q-\psi  \) with a concavity
change, would give similar results with different values for the \( u \)
depending functions in the relations (\ref{jet},\ref{shear},\ref{ym})
. This may be a way of improving these results.

However, our feeling is that such a search for improvement is of little
interest, given that the 1-1/2 Quasi-Geostrophic is a crude model
of the troposphere of Jupiter. Firstly, the geostrophic balanced is
not well verified in the area where the curvature of the jet is minimal,
and the layer height variations are not very small compared to the
layer height. A 1-1/2 Shallow-Water model would improve the results,
and be more convenient to make a very precise study. Secondly, a 1-1/2
layer is certainly a crude approximation of the Jovian troposphere.

\subsection{Validity of the lower order approximation}

The model we propose assumes several hypothesis (PV distribution,
optimal mixing) and approximations (for instance the Quasi-Geostrophic
model 1-1/2 model). This numerical computation allows to test the
approximation made when describing the solution by its lower order
description when \( R\rightarrow 0 \). For instance, for the width
of the jet, we have obtained \( l^{\star }_{x}=5100\, km \) and \( l^{\star }_{y}=4100\, km \).
The jet width thus depends on latitude (this is visible on both the
computed and the observation velocity fields, figure \ref{fig:Emergence_Numerique_tache_rouge}\textbf{)}
and is larger than the first order prediction. This is a consequence
of the shear (for \( l^{\star }_{y} \)) and of the strong curvature
of the jet near the extrema of the topography (for \( l^{\star }_{x} \)).
There, the curvature \( r \) is only 3 times the Rossby deformation
radius \( R \), which limits the validity of the approximation \( R\rightarrow 0 \).
We note that the latitudinal dependence is present at the following
order of the asymptotic expansion (see Bouchet 2001, part 1, section
4.3). 

We also note that the curvature at mid latitude is greater for the
numerically computed equilibrium than the analytical prediction. This
is also due to the small \( R \) approximation. This curvature difference
in this area is important, however this has only a limited effect
on the maximal latitudinal extension of the spot. Indeed, the value
\( y_{m} \) analytically predicted is \( y_{m}=4300\, km \), whereas
the numerically computed one is \( y_{m}=3800\, km \). This explain
why the shape of the spot is correctly predicted by the analytical
relation, in spite of the limitations of the lower order approximation.
The curvature of the real jet is also smaller than the one of the
numerically computed one. This may be due to the discrepancies of
the Quasi-Geostrophic approximation in this area (cyclostrophic balance).

We thus conclude that the lower order of the small Rossby deformation
radius is valid only as a first approximation, for a value of \( R \)
corresponding to the GRS (let say \( 30\% \) for the described variables).
However, we note that the qualitative agreement is very good. In particular,
the prediction of the latitudinal maximal extension, and of the vortex
shapes is good. Whereas the numerical values are not exact, this analysis
has permitted us to understand the role of the various parameters,
in order to find parameters suited to model the GRS, the White Ovals
(section \ref{sec:Emergence_Num_OBC}) or the Brown Barges (section
\ref{sec:Emergence_Num_Jets}).

\section{Thermodynamic phase transition and strong jet stability.}

\label{sec:Emergence_Num_Jets}

Because of the very different typical time scales, for forcing and
dissipation in on hand, and for inertial organization in the other
hand, Jupiter's features appear stationary. For instance, the Great
Red Spot is observed from more than three centuries. Whereas its length
seems to have change during this time, its global structure is likely
to be the same. Jovian feature should therefore be stationary for
the inertial dynamics. 

A large amount of work has dealt with the stability of quasi-two dimensional
flows. Linear stability of jets has been for instance addressed by
Rayleigh, Kuo, Charley and Stern (see Pedlosky (1987) for a discussion
for geophysical flows). Nonlinear stability results have first be
obtained by Arnold (1966) for the Euler equation. The flows are then
proven to be stable because they minimize a functional, built on the
Casimirs and on the energy, invariant under the dynamics (formal stability).
A further estimate on this functional allows to prove that a perturbation
around the stationary state remains bounded under the nonlinear dynamics
(non-linear stability). A generalization of these ideas for other
flow equations have then been studied (see for instance Holm and collaborators
(1985) or Yongming, Mu and Sheperd (1996) for geophysical flows).
In the case of zonal solutions, for the Quasi-Geostrophic equation
or for the Euler equation, the linear stability results can be retrieve
from the nonlinear stability results.

These stability results are only sufficient condition for stability.
Lots of geophysical flows, essentially the most energetic ones, are
indeed stable whereas they do not verify the hypothesis of these theorems.
For the Jovian atmosphere, this is for instance reported in the review
of Dowling (1995). This has led to some interrogation on the stability
of these flows. These questions have been emphasized also by the difficulty
to obtain numerical model of such flows, having strong jets, typical
of the Jovian troposphere. 

The statistical mechanics of the potential vorticity offers a way
to understand this stability. The link between the Arnold's stability
theorems and the statistical equilibrium has been noted for instance
by Robert and Sommeria (1991) (see also Bouchet (2001)). In such works
the equivalent of the Arnold's theorem hypothesis is that only one
solution exist for a given inverse temperature \( \beta  \). This
has been proven for states with \( \beta >\beta _{c} \) (or equivalently
for energy sufficiently small \( E<E_{c} \), as proven in section
\ref{sec:zonal_topographie}). For smaller \( \beta  \) (larger energy
\( E>E_{c} \)), the stability of the flow was qualitatively understood
by the impossibility of the potential vorticity to mix further. A
clear formalization of this statement has been proposed recently by
Ellis, Haven and Turkington (2002), where an augmented functional,
taking into account of the Energy conservation, have been used to
generalize the Arnold's stability theorem. The result of this work
is that any entropy maxima under constraints, except the ones close
to a phase transition point, is stable. As no norm is specified in
this work, and the a-priori estimate necessary to prove a nonlinear
stability theorem is not provided, these results are the proof of
formal stability (see Holm and collaborators (1985) for a precise
definition of formal stability) of such flows. This is however a decisive
step towards the understanding of the stability.

A crucial hypothesis needed to use these results is that the second
variations of the augmented functional used by Ellis, Haven and Turkington
(2002), be definite positive. In Ellis, Haven and Turkington (2002),
this point is not analyzed in details, and cited as a technical problem.
Unfortunately, this is not right in most of the situations. As soon
as the equations have some symmetry, and the flow does not respect
this symmetry, the second variations can not be definite positive.
At least one direction must have a zero eigenvalue. For instance,
in our case, the vortex solutions break the zonal symmetry. A small
perturbation of such a vortex, can lead to the translation of the
vortex by a finite distance. For this reason, a nonlinear stability
result is not possible. However, the situation is physically very
clear: a perturbation can only lead to flows close to the initial
ones, up to an arbitrary translation. A clear formalization of these
ideas remains to be done. Anyway, the results of Ellis, Haven and
Turkington (2002) are a decisive step towards the understanding of
the stability of such flows.\\

\begin{figure}

\caption{Phase diagram of the Gibbs states versus the energy \protect\( E\protect \)
and the asymmetry parameter \protect\( B\protect \), with a quadratic
topography and a domain aspect ratio corresponding to the Great Red
Spot parameters. (400.000 km over 20.000 km). The outer line is the
maximum energy achievable for a fixed \protect\( B\protect \) : \protect\( E=\frac{R^{2}}{2}(1-B^{2})+\mathcal{O}\left( R^{3}\right) \protect \).
The inner solid line corresponds to the frontier between the vortex
and straight jet solutions. The dash line corresponds to the limit
of validity of the small deformation radius hypothesis. It has been
drawn using the condition that the maximal vortex width (\ref{ymax})
is equal to two Rossby deformation radius. The dot lines are constant
vortex aspect ratio lines with values 2,10,20,30,40,50,70,80 respectively.
We have represented only solutions for which anticyclonic PV dominate
(\protect\( B>0\protect \)). The opposite situation may be recovered
by symmetry.}

\label{phase_top}

\resizebox*{0.8\textwidth}{!}{\includegraphics{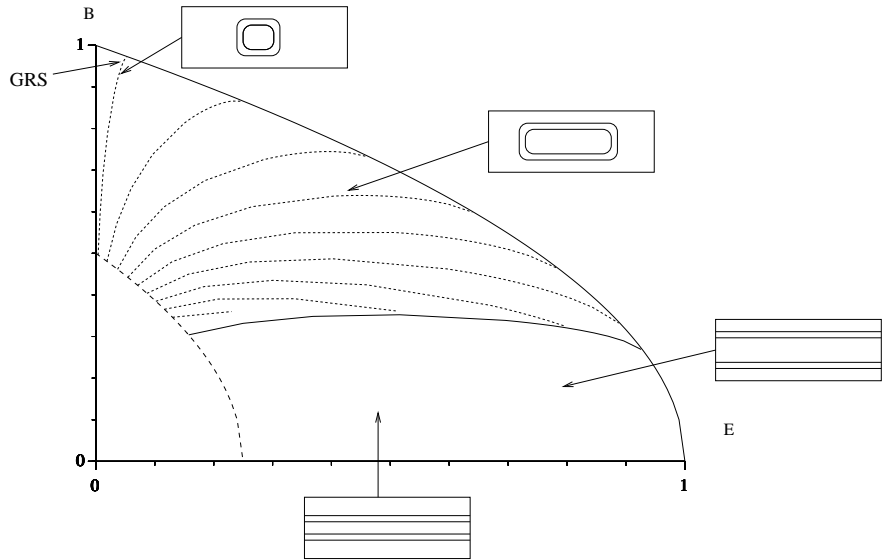}}  
\end{figure}

Statistical equilibrium are thus stable, as soon as they are not too
close to a phase transition. On a practical point of view, one thus
have to study the phase diagram of the equilibrium states. For instance
phase diagrams on figure \ref{phase} and \ref{phase_top} represent
stable stationary flows, similar to Jovian vortices and jets. Please
note that some other stable states may exist for the same energy and
parameter \( B \) (metastable states, for instance).

As illustrated on figure \ref{phase_top}, depending on the parameters,
strong jets or vortex solutions may be stable, depending on the values
of the parameters \( E \) and \( B \). In order to illustrate these
stability properties, we consider the evolution of an initial condition
composed of an anticyclonic PV band, standing on the maximum of the
topography. The corresponding flow is made of two strong zonal jets
flowing eastward and westward respectively. The PV has a width \( l \).
We strongly perturb this initial condition by centering the PV band
on a latitude \( y_{c} \) varying with the longitude: \( y_{centre}=\pi /2+l/4\sin x+l/12\sin (3\pi /2+\pi /6) \).
Figures \ref{fig:Emergence_Numerique_Jet} and \ref{fig:Emergence_Numerique_Jet_Tache}
show this initial condition, for \( l=1.09 \) and \( l=0.31 \) respectively.
The values of the topography parameter (see \ref{Topographie}) and
of the Rossby deformation radius are \( a=0.6 \) and \( R=0.25 \)
respectively. 

The parameters for the numerical simulation are \( \nu =5.93\, 10^{-5} \)
(resolution 512X128), corresponding to a Reynolds number (based on
the Rossby deformation radius) of \( Re=\left( Rv_{max}\right) /\nu  \)
equal to 820. The numerical time step is \( \Delta t=0.012 \). 

Figure \ref{fig:Emergence_Numerique_Jet} illustrates the evolution
for the first initial conditions. The first picture show that the
two jets, are destabilized by this strong perturbation. The jet however
stabilize accordingly to the phase diagram on figure \ref{phase_top}
(the value of \( B \) is then close to zero). We note that this initial
condition does not verify the non-linear or linear stability theorem
hypothesis. The last of these pictures show slight oscillations of
the PV level lines, that we interpret as Rossby waves, guided by the
jet. The relaxation of these waves is very slow. 

Figure \ref{fig:Emergence_Numerique_Jet_Tache} illustrates the evolution,
for the second initial condition. As the area of the PV band is then
small, the value of \( B \) is no more close to \( 0 \). Accordingly
to the phase diagram on figure \ref{phase_top}, the two jet destabilize
and form anticyclones. The statistical equilibrium is then an elongated
anticyclone centered, on the topography extrema. The final state is
shown on figure \ref{fig_Brown_Barges}. This solution will also be
used in next section to model one of the cyclonic Brown Barges of
the Jupiter's north hemisphere.

\section{The north-hemisphere Brown Barges}

\label{sec_Brown_Barges}

Brown Barges are brown oval spots (see figure \ref{fig_ellipse}),
located at \( 14^{o} \) N on the Jupiter's troposphere. On the contrary
to most of Jovian features, these vortices are cyclones. A study of
these spot velocity field is reported in Hatzes and collaborators
(1981). In this section, we model this spot with the statistical equilibrium
vortex obtained in the previous section, from the destabilization
of a strong jet (figure \ref{fig:Emergence_Numerique_Jet_Tache}).
We compare the velocity field of this statistical equilibrium structure
with the data analysis of Hatzes and collaborators (1981). 

The equilibrium PV field obtained from this numerical simulation is
represented on figure \ref{fig_Brown_Barges}. This figure actually
represents an anticyclone. However, because of the symmetries of the
Quasi-Geostrophic model, a symmetric cyclonic structure may be obtained.
The very elongated shape, with a maximum latitudinal extension may
be compared to the image of one of the real Brown Barges (see figure
\ref{fig_ellipse}). We also represent the velocity in a latitudinal
and meridional sections of the spot, both for the statistical equilibrium
and for the data analysis of Hatzes and collaborators (1981). One
may observe that the velocity structure is the same. The northward
velocity, along a meridional section, shows a strong jet structure
at the front edge of the spot, whereas it is null inside of the spot.
On the contrary, the eastward velocity along a zonal section, does
not show the jet structure : it is essential a shear flow. The comparison
of these plots shows that statistical equilibrium describe very well
the qualitative properties of this spot. We have not tried to give
some dimensional values of the main characteristic of the spot. However
this could be done. By an iterative algorithm, one could then try
to predict the actual values of the topography curvature and of the
Rossby deformation radius, as we have done for the Great Red Spot.

The structure is thus different from the Great Red Spot one's. The
jet structure on the northward velocity allows to conclude that the
Rossby deformation radius is small. The effect of a very intense topography
curvature characterizes the Brown Barges. For this reason, the topography
can no more be treated as a first order perturbation, like has been
done in section (\ref{sec:Equilibrium_Small_Rossby}). However, an
asymptotic description, following the same ideas can be done, assuming
the amplitude of the topography as having effects on the zeroth order
of the asymptotic expansion. The result is a modified algebraic equation
describing the velocity field outside of the spot. The jet structure
then explicitly depends on the latitude \( y \). The shape of the
vortex is always described by an equation similar to (\ref{Rayon_courbure}).
However, the dependence on \( y \) because of the topography, is
no more due only to the PV variation inside of the spot, but also
to the latitudinal dependence of the jet properties. We leave a more
precise description and study of this asymptotic expansion for future
works. 

In Hatzes and collaborators (1981), authors insist on the oscillations
of the shape of this cyclone, whereas we have described only a stationary
solution, with similar velocity field structure. We note that, as
in the case of the jets described in section \ref{sec:Emergence_Num_Jets},
perturbation of the equilibrium structure would lead to oscillations
around the stationary flow which should describe the observed ones.
A further study of this problem may be of interest.

\section{Conclusion}

We have described equilibrium of the potential vorticity statistical
equilibrium, for the Quasi-Geostrophic equation. Independently of
the statistical interpretation, all the flow we have described are
stable stationary flow for the inviscid dynamics. We have first presented
results in the limit of small Rossby deformation radius. The main
ideas of this asymptotic description are present in the work Bouchet
and Sommeria (2002). The derivation proposed here is however simplified.
We have discussed in details the generalization of these results to
any potential vorticity distributions. Using numerical computations,
we have also described statistical equilibrium flows for parameters
for which the hypothesis of a small Rossby deformation radius no more
holds. The application of these results to model, the Jovian strong
jets and main vortices are extensively described.

In the limit of small Rossby deformation radius, the equilibrium flows
are characterized by strong jets. The minimization of the entropy
selects either zonal solutions or vortices bounded by strong jets,
depending on the parameters. These jets play the role of an interface
separating two area of different potential vorticity mixing. The shape
of this interface is given by the minimization of their length, for
a given area. Under the presence of a deep zonal flow and of a beta-effect,
or equivalently of a topography, this minimization is balanced by
the tendency of anticyclonic potential vorticity to stand around the
maxima of the topography. This leads to the characteristic elongated
vortices observed on Jupiter's troposphere. The width of these vortices
may be computed exactly. A typical width is given by an alternative
of the Rhine's scale, built on the curvature of the topography, or
equivalently on the deep layer shear \( \sigma _{d} \) : \( L=\left( RU/\sigma _{d}\right) ^{1/3}=R\left( \Omega /\sigma _{d}\right) ^{1/3} \).
This model predicts that vortices sit on extrema of the topography.
This property has been verified using available data for the Great
Red Spot and the White Oval BC.

Using these results, we have proposed a quantitative model for the
Great Red Spot. In Bouchet and Sommeria (2002), using the small Rossby
radius derivation, with a two PV-levels distribution, we have shown
that an appropriate choice of the energy, total potential vorticity,
topography curvature and Rossby deformation radius allows to reproduce
the observed jet maximum velocity, jet width, vortex shape and aspect
ratio, and surrounding shear. In this work, by comparison with numerical
computation of the equilibrium, we have shown that the small Rossby
radius approximation leads to a correct description of the structure
of the vortex, for the Great Red Spot parameters. The discrepancies
due to finite size effects are of the order of the error due to the
Quasi-Geostrophic approximation. The obtained velocity field compares
very well with the observed one. 

Using numerical determination of the equilibrium flows, we have proposed
models of the White Ovals or of the cyclonic Brown Barges. These vortices
may be obtained either from random initial conditions or from the
destabilization of strong jets. The White Ovals are characterized
by a size which is of the same order as the Rossby deformation radius.
The deep shear is responsible for their oval shape and for the surrounding
shear. The Brown Barges are characterized by a very strong topography
curvature. For these vortices, we have obtained their typical jet
like structure for the northward velocity in a zonal section, and
their typical shear for the eastward velocity in a meridional section.

The statistical mechanics predicts some strong qualitative properties
for Jovian like vortices. For a given deep shear, and Rossby deformation
radius (for the same latitudinal band, for instance), smaller vortices
are close to a circle. Their aspect ratio grows with the size. The
latitudinal extension of the spot as a maximum value, forcing a very
elongated shape as the one of the Brown Barges. For such vortices
to exist, a critical energy in the latitudinal band must be present.
When this is the case, the shear in the active layer has to be larger
than the shear in the deep layer. Vortex adapt their zonal drift speed,
such as in their reference frame, they are located in an extrema of
the topography. Thus similar vortices in the same latitudinal band
must have a relative zonal velocity if they are not located at the
same latitude. This drift velocity is linear with the latitude difference,
as soon as the latitude difference is sufficiently small for the deep
shear to be supposed constant.

All the statistical equilibrium are dynamically stable, even when
the conditions for linear or non-linear classical results do not apply.\\

The statistical mechanics of the potential vorticity describes the
most probable flow for a given energy and potential vorticity distribution.
The main assumption of this work is that such flows describes actually
the observed jets and vortices of Jupiter's troposphere. The dynamical
system ergodicity would be a sufficient condition to justify this
hypothesis.  The proof of ergodicity is very difficult, even for very
simple dynamical systems. The best way to study the validity of statistical
mechanics in complex systems remains the comparison of its predictions
with observations of with numerical simulations. This works has shown
that the statistical mechanics of the potential vorticity is able
to model the main Jovian features and to predict important qualitative
properties, which can be verified. 

In a latitudinal band, the topography forces a shear flow in the active
layer. This favors vortex merging, and thus the potential vorticity
mixing. On the other hand, if we consider a topography with two bumps,
following the same ideas as in this work, we may for instance describe
stable solutions corresponding to two anticyclones on each of the
topography bumps. In these two bumps are of different high, this solution
may not be a statistical equilibrium state, but only a local maxima
of the entropy under constraint. In such a case, the topography would
act as a dynamical barrier, preventing a real ergodicity for the system.
However, the statistical interpretations would still remain clear. 

An other point to be discussed is the slowing down of the dynamics,
due to the small value of the Rossby deformation radius. The interactions
between vortices, decays exponentially for distance much larger than
\( \textrm{R} \). For this reason, well separated vortices mainly
not interact. This may prevent their merging predicted by statistical
mechanics. It will at least impose a very long time scale to observe
an effective ergodicity. This situation is illustrated by the merging
of the three White Ovals which formed in 1938 (Sanchez-Lavega and
collaborators 2000). The two last ones have finally merged in 2002,
after a very long coexistence in the same latitudinal band.

Besides these qualitative arguments, we want to stress that ergodicity
may really be questionable in some situations, for such systems. As
an example we refer to Barré and collaborators (2002). In this study,
a system with long range interactions, sharing deep analogies with
quasi-two-dimensional flows, is shown to have very long lived out-of
equilibrium states.\\

We have discussed some limitations of these statistical models for
Jupiter's jets and vortices. The major ones concerns the validity
of the modeling of Jupiter's troposphere by the Quasi-Geostrophic
model. Whereas it is good as a first approximation, generalization
of these results for a Shallow Water dynamics or for multi-layered
dynamical models should provide more precise results in order to propose
more precise comparison with the observed structures.

Some further studies of the equilibrium structure should be of interest
to put precise constraint on the physical parameters. For instance,
it may be possible to determine the deep shear under major vortices,
such as the White Ovals or the Brown Barges. A model of White Ovals
merging based on statistical mechanics could also permit to put further
constraints on the physical parameters and potential vorticity distribution.
Studies of linear perturbations around the equilibrium structures
should be able to describe the spot shape oscillations, as observed
for instance for the Brown Barges (Hatzes and collaborators, 1981). 

In this work, we have assumed an inviscid dynamics. This is a very
natural assumption, given the very different time scales for forcing
and dissipation in one hand, and inviscid organization in the over
hand. Whereas the observed features should be inviscid statistical
equilibrium, the actual PV distribution and energy actually depends
on the forcing and on the dissipation. We have for instance argued
for a distribution with bounded potential vorticity, which should
be well approximated by a two level distribution, because of the observed
forcing by incoming thermal plumes. However, in order to go further
in the analysis of the potential vorticity fine-grained distribution,
one should model more precisely forcing and dissipation. This could
permit to explain observable phenomena, such as the diminution, on
a very long time scale, of the size of the giant anticyclones, or
an eventual inviscid evolution of the oscillations of some vortices.
In practice, the forcing should be introduced in kinetic like equations,
like those developed in Kazantzev and collaborators, or following
ideas described in DiBattista, Majda and Grote (2001). Such models
would explain the long term evolution of the fine-grained PV-distribution,
whereas the observed structure should remain close to equilibrium
structure during the evolution.

\section*{Acknowledgments}

This work is part on the PHD thesis of F. Bouchet, directed by R.
Robert and J. Sommeria. The first part of this work (Bouchet and Sommeria
2002), has been done in collaboration with J. Sommeria. The authors
thank J. Sommeria, R. Robert and P.H. Chavanis for useful comments
on the present work. During this work, one of us has been funded by
a {}``Bourse Lavoisier'' of the French {}``Ministère des Affaires
Etrangères'', and by the program COFIN00 {}``chaos and localization''.

\appendix

\section{Analysis of the jet shape equation }

\label{appendice_forme_jet}In section \ref{sec:Equilibrium_Small_Rossby},
we have obtain the equation verified by the jet position (\ref{Rayon_courbure}):
\( \epsilon u\left( h(y)-\alpha _{1}\right) =\frac{e(u)}{r} \). In
this section, we discuss equations allowing numerical integration
of this equation and we derive an analytical expression for the latitudinal
extension of a vortex solution. This derivation, is a generalization
to a cosine topography, of results in Bouchet and Sommeria (2002).

We look at anticyclones solutions (\( \varepsilon =1 \)) around the
maxima of \( h \) (\ref{Topographie}). This extrema is reached for
\( y=\pi /2 \). We make a latitudinal translation such that the maxima
of the topography be on \( y=0 \). We thus consider the topography
\( h\left( y\right) =2a\cos \left( 2y\right)  \). The cyclone case
may be easily recovered by symmetry. 

To make the equation for the radius of curvature more explicit, let
us define \( s \) a curvilinear parameterization of our curve, \( {\textbf {T}}(s) \)
the tangent unit vector to the curve and \( \theta (s) \) the angular
function of the curve defined by \( {\textbf {T}}(s)=(\cos \theta (s),\sin \theta (s)) \)
for any \( s \). Then the radius of curvature \( r \) of the curve
is linked to \( \theta (s) \) by \( 1/r=d\theta /ds \). This yields
the differential equations : \begin{eqnarray}
\frac{d\theta }{ds} & = & \frac{2au}{e\left( u\right) }\cos \left( 2y\right) -\alpha _{1}\label{tht} \\
\frac{dy}{ds} & = & \sin \theta (s)\label{thy} \\
\frac{dx}{ds} & = & \cos \theta (s)\label{thx} 
\end{eqnarray}
For symmetry reasons, it is easily verified that the solutions of
(\ref{tht}, \ref{thx} and \ref{thy}), with initial conditions \( \theta (0)=\frac{\pi }{2} \),
\( y(0)=0 \) and some \( x(0) \) are periodic. The resulting vortex
is then symmetric with respect to the latitude of the maxima of the
topography (here \( y=0 \)). Moreover, we have proved in the appendix
C of Bouchet and Sommeria (2002) that these initial conditions are
the only ones leading to vortex solutions (closed curves), in the
case of a quadratic topography. The argument used there only uses
the symmetry of the topography with respect to the extrema of the
topography and can easily be generalized in the present case.

Let us compute \( y_{m} \), the maximal latitude of the vortex (the
maximal latitude of the jet center) (\( y_{m} \) is the half of the
latitudinal extension of the vortex). We first note that the two variables
\( \theta  \) and \( y \) are independent of \( x \). We will therefore
consider the system formed by the two first differential equations
(\ref{tht},\ref{thy}). It is easily verified that this system is
Hamiltonian, with \( \theta  \) and \( y \) the two conjugate variables
and \begin{equation}
\label{H}
H\equiv \cos \theta +\frac{au}{e\left( u\right) }\sin \left( 2y\right) -\alpha _{1}y
\end{equation}
 the Hamiltonian. Thus \( H \) is constant on the solution curves.
From the initial condition, we deduce \( H=1 \). We note that for
\( y=y_{m} \), the curvature of the jet is \( 0 \) (\( 1/r=0 \)
). From \ref{tht}, we thus obtain \( \frac{2au}{e\left( u\right) }\cos \left( 2y_{m}\right) -\alpha _{1}=0 \).
Combining this relation with the one obtained by using that \( H=1 \),
for \( y=y_{m} \) and \( \theta =\pi  \) : \( -1+\frac{au}{e\left( u\right) }\sin \left( 2y_{m}\right) -\alpha _{1}y_{m}=1 \)
allows to compute \( y_{m} \) and \( \alpha _{1} \). This gives:\[
y_{m}=\frac{1}{2}g\left( \frac{e\left( u\right) }{2au}\right) \, \, \, \, and\, \, \, \, \alpha _{1}=\frac{2au}{e\left( u\right) }\cos \left( 2y_{m}\right) \]
 where \( g \) is the inverse of the function \( \sin x-x\cos x \)
for \( 0<x<\pi  \).

\newpage

\textbf{Figure Captions.}\\

\newpage

\newpage

\newpage

newpage

\begin{figure}
{\centering \resizebox*{!}{0.8\textheight}{\includegraphics{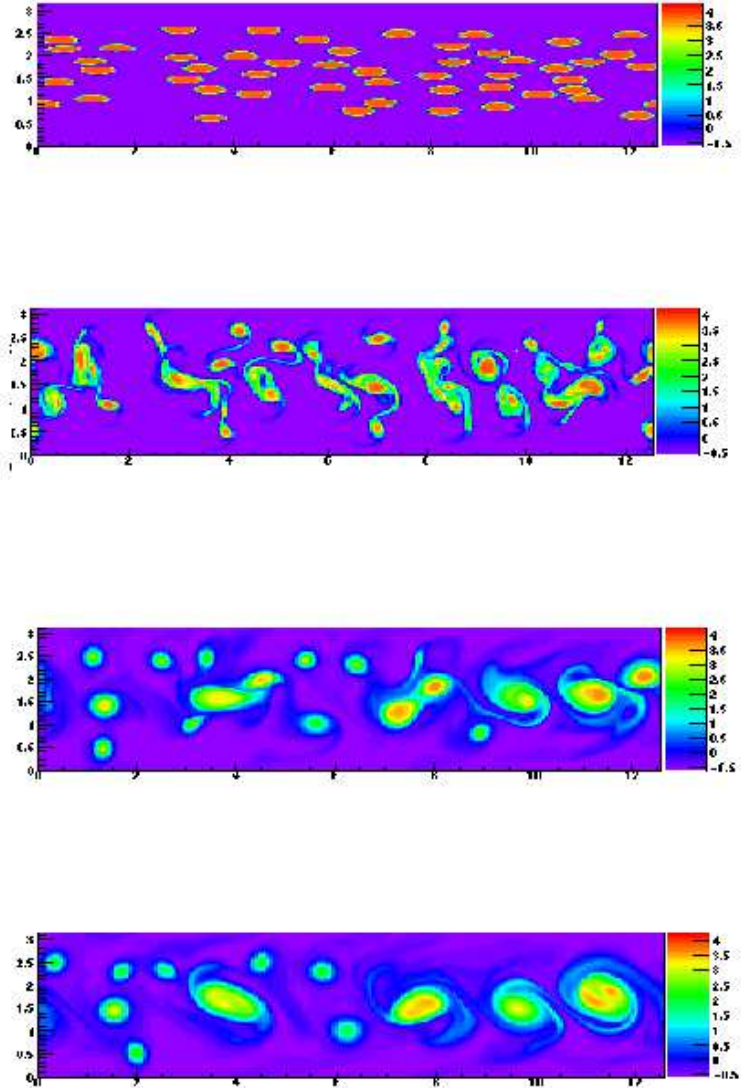}} \par}

\caption{Dynamics of random initial vorticity patches, in the Quasi-Geostrophic
model, using a small scale parameterization based on a maximum entropy
production principle (\foreignlanguage{english}{\ref{Emergence_Num_relaxation}}).
The color represents the PV values. The Rossby deformation radius
is very small (\protect\( R=0.2\protect \)), compared to the latitudinal
band width (\protect\( \pi \protect \)). We use a cosine topography
(\ref{Topographie}) whose maxima is located at the center of the
latitudinal band. The later evolution is shown on figures \ref{fig:Emergence_Numerique_OBC2}
and \ref{fig:Emergence_Numerique_OBC3}\label{fig:Emergence_Numerique_OBC1} }
\selectlanguage{english}
\end{figure}

\begin{figure}
{\centering \resizebox*{1\textwidth}{!}{\includegraphics{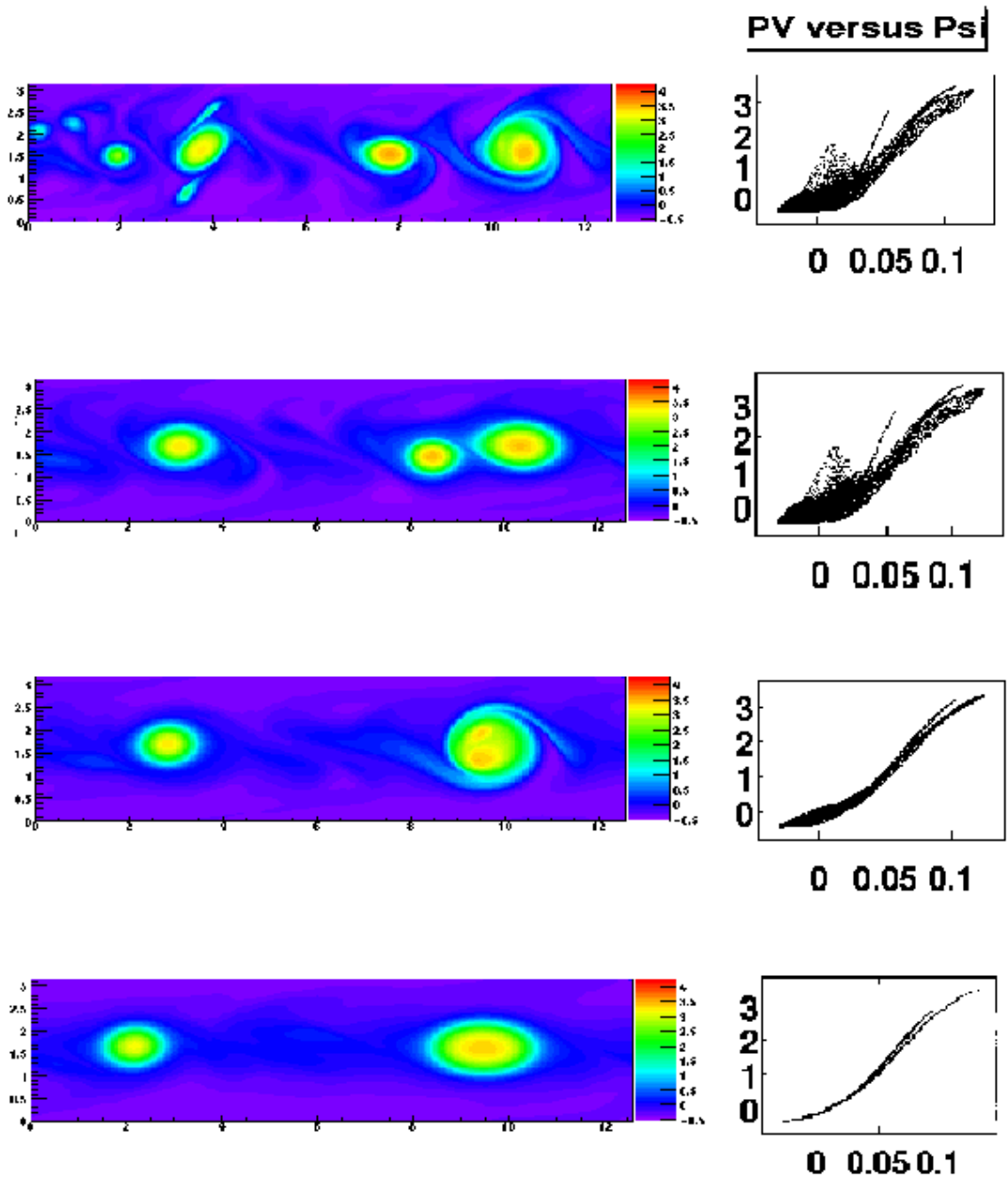}} \par}

\caption{Continuation of the previous figure. After a very rapid local organization,
three anticyclones form. On a much longer time scale, they merge,
forming elongated vortices similar to the White Ovals on the jovian
troposphere (the time lapse between the scatter-plots of figure \ref{fig:Emergence_Numerique_OBC1}
is approximately 16 turn over times, whereas between the two last
scatter plots of this figure it is 50 turn over times, and 300 to
obtain the final organization represented on figure \ref{fig:Emergence_Numerique_OBC3}.
The insets show Stream function-PV scatter plots. They illustrate
the evolution towards stationary states.\label{fig:Emergence_Numerique_OBC2} }
\selectlanguage{english}
\end{figure}

\begin{figure}
{\centering \resizebox*{1.1\textwidth}{!}{\includegraphics{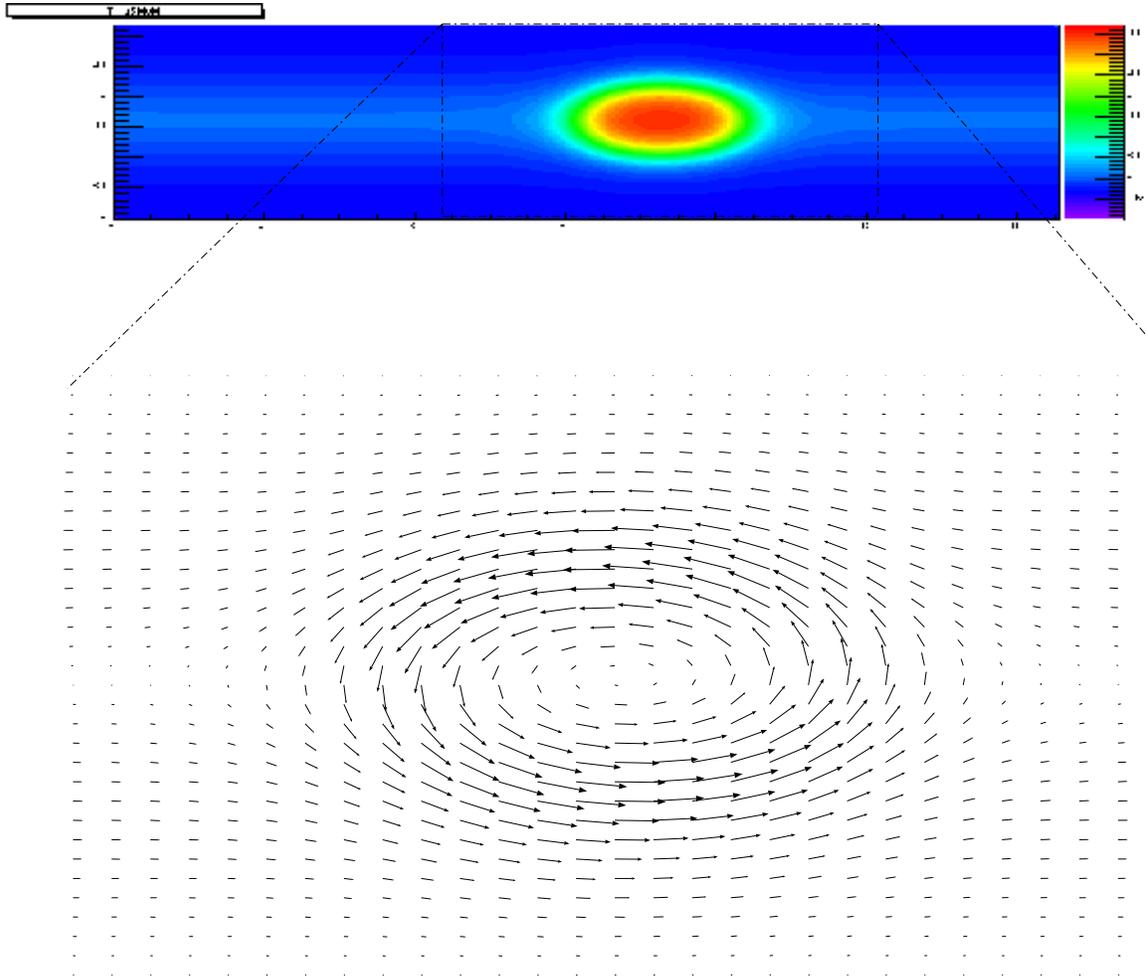}} \par}

\caption{Equilibrium structures corresponding to the dynamical evolution of
the two preceding figures. The upper figure is the PV, whereas the
lower one is the velocity field. The maxima of entropy under constraint
is an anticyclone, centered on the maxima of the topography. The surrounding
shears and its oval shape are consequences of its interaction with
the deep layer flow (topography). This structure is similar to the
one of the White Ovals. It differs from the Great Red Spot, because
the Rossby deformation radius is of the same order as the vortex size.
\label{fig:Emergence_Numerique_OBC3} }
\selectlanguage{english}
\end{figure}

\begin{figure}
{\centering \resizebox*{1.1\textwidth}{!}{\includegraphics{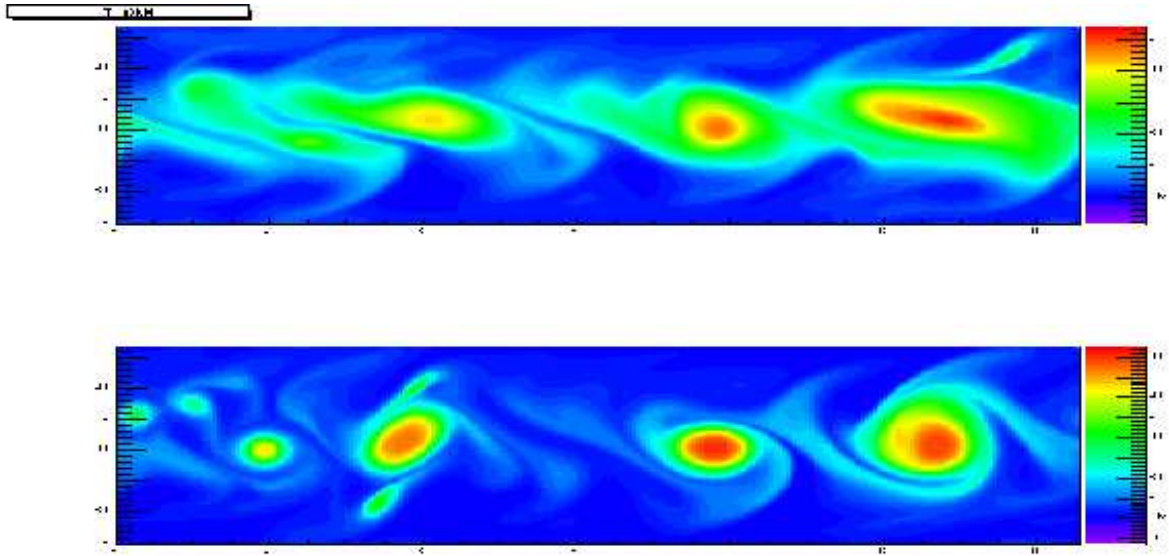}} \par}

\caption{Comparison of the evolution of the same initial condition, made of
random vorticity patches (the same as for the three previous figures),
for two different small scale Potential Vorticity mixing parameterization.
The upper figure shows the result for a Direct Numerical Simulation
(usual eddy diffusivity), the upper one shows the results for the
relaxation equations (\ref{Emergence_Num_relaxation}). This shows
that the Direct Numerical Simulation does not allowed to obtain strong
coherent vortices, for very long times.\label{fig:Emergence_Numerique_Comparaison_relaxation_NS} }
\selectlanguage{english}
\end{figure}
 
\begin{figure}
{\centering \resizebox*{1.1\textwidth}{!}{\includegraphics{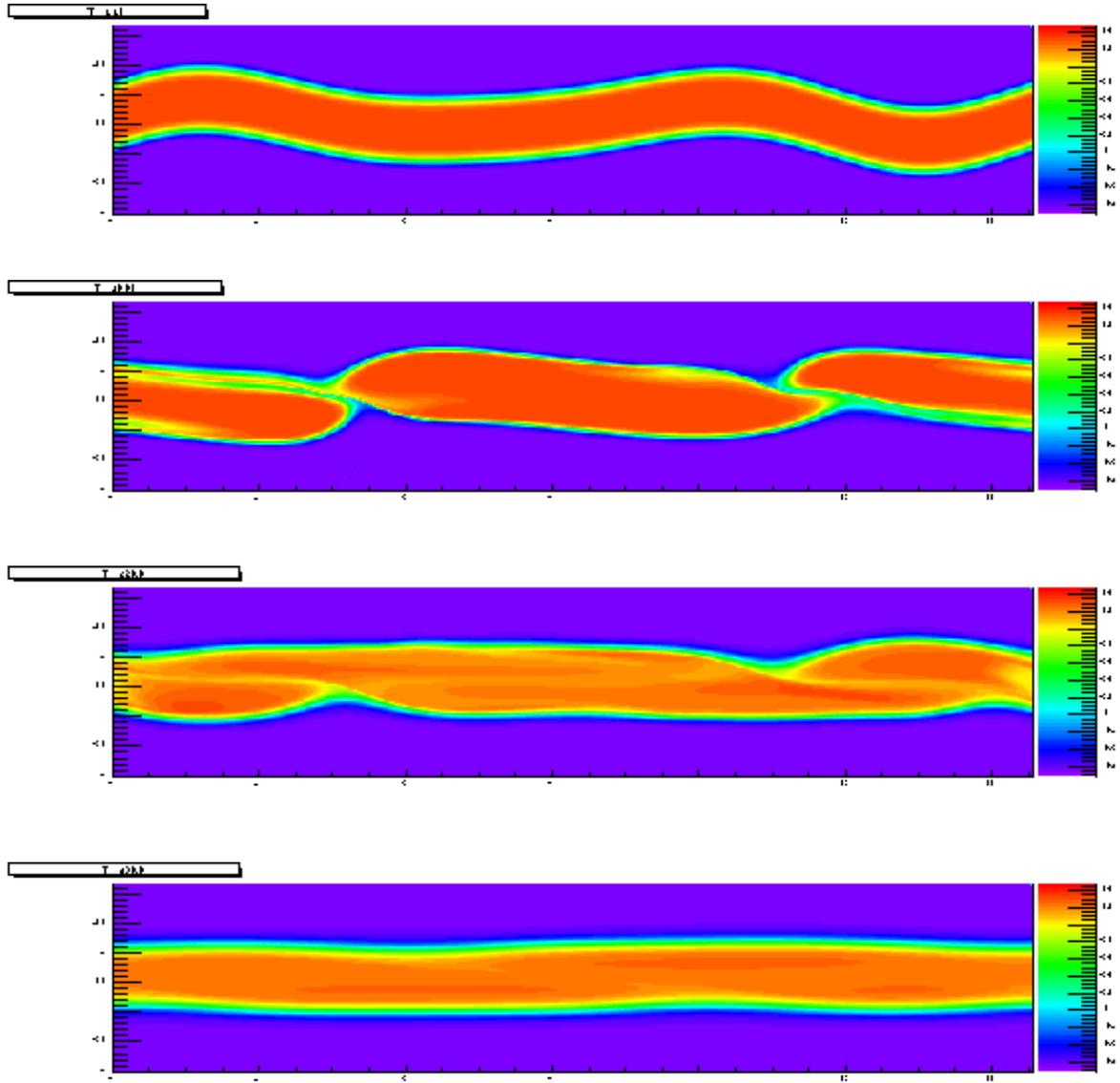}} \par}

\caption{Potential Vorticity field, for two jets flowing eastward and westward
respectively. Whereas they do not verify the two Arnold's theorem
hypothesis, submitted to a strong perturbation, they stabilize. The
maximization of the entropy under constraint allow to obtain new stability
theorems (see section \ref{sec:Emergence_Num_Jets}). \label{fig:Emergence_Numerique_Jet} }
\selectlanguage{english}
\end{figure}

\begin{figure}
{\centering \resizebox*{1.1\textwidth}{!}{\includegraphics{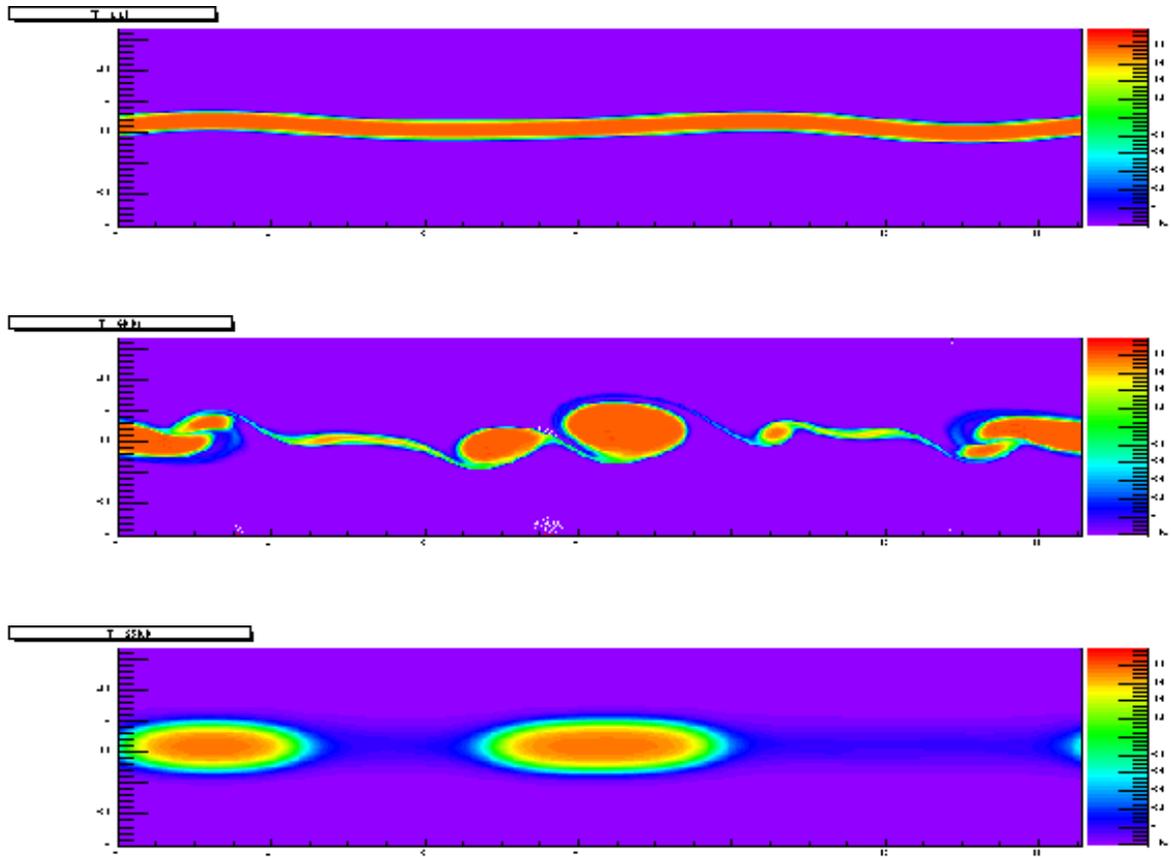}} \par}

\caption{Destabilization of two strong jets, and formation of very elongated
vortices similar to the cyclonic Brown Barges in the north hemisphere
of the Jovian atmosphere. The stability property of such jets and
vortices is summarized by the phase diagram on figure \ref{phase_top}
(see section \ref{sec:Emergence_Num_Jets}). \label{fig:Emergence_Numerique_Jet_Tache} }
\selectlanguage{english}
\end{figure}

\begin{figure}

\caption{\label{fig_Brown_Barges}The upper figure shows the Potential Vorticity
field for a statistical equilibrium on a strong topography. The shape
of the spot can be compared to real image of the Brown Barges (figure
\ref{fig_ellipse}). The four lower figures show the velocity for
a zonal section (eastward velocity, left figures) and for a meridional
section (northward velocity, right figures). The two upper velocity
figures are the observed ones for one of the cyclonic Brown Barges,
in the north hemisphere of Jupiter (from Hatzes and collaborators
(1981)). The two lower ones are the statistical equilibrium ones. }

{\centering \begin{tabular}{c}
\resizebox*{!}{0.2\textheight}{\includegraphics{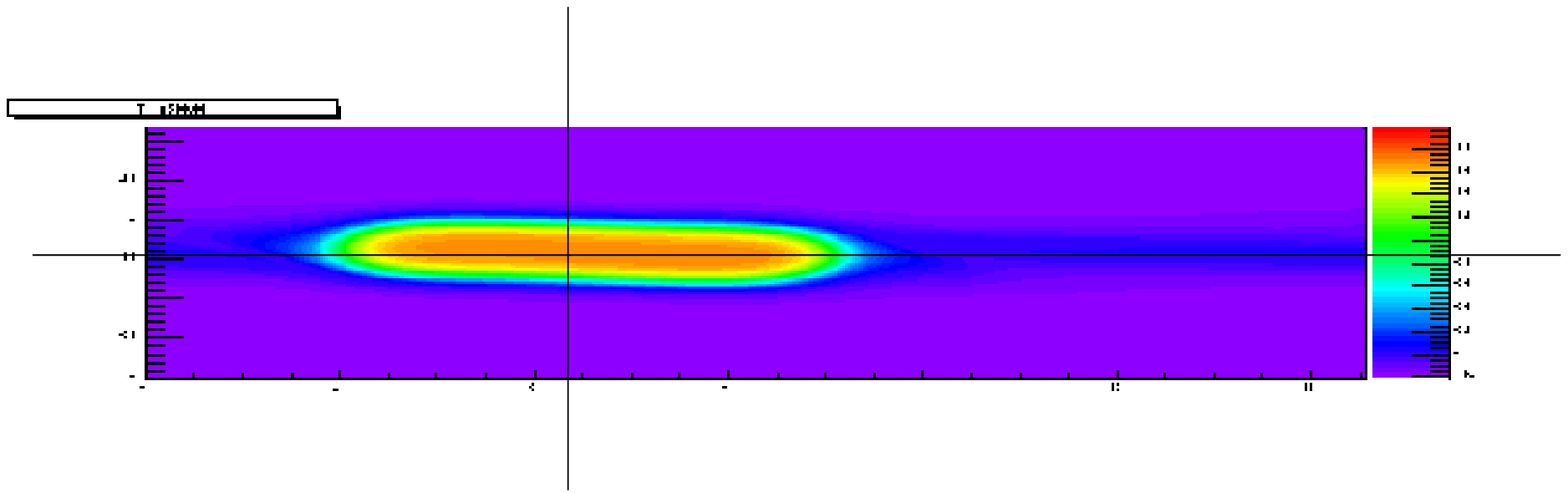}} \\
\begin{tabular}{cc}
\resizebox*{!}{0.4\textheight}{\includegraphics{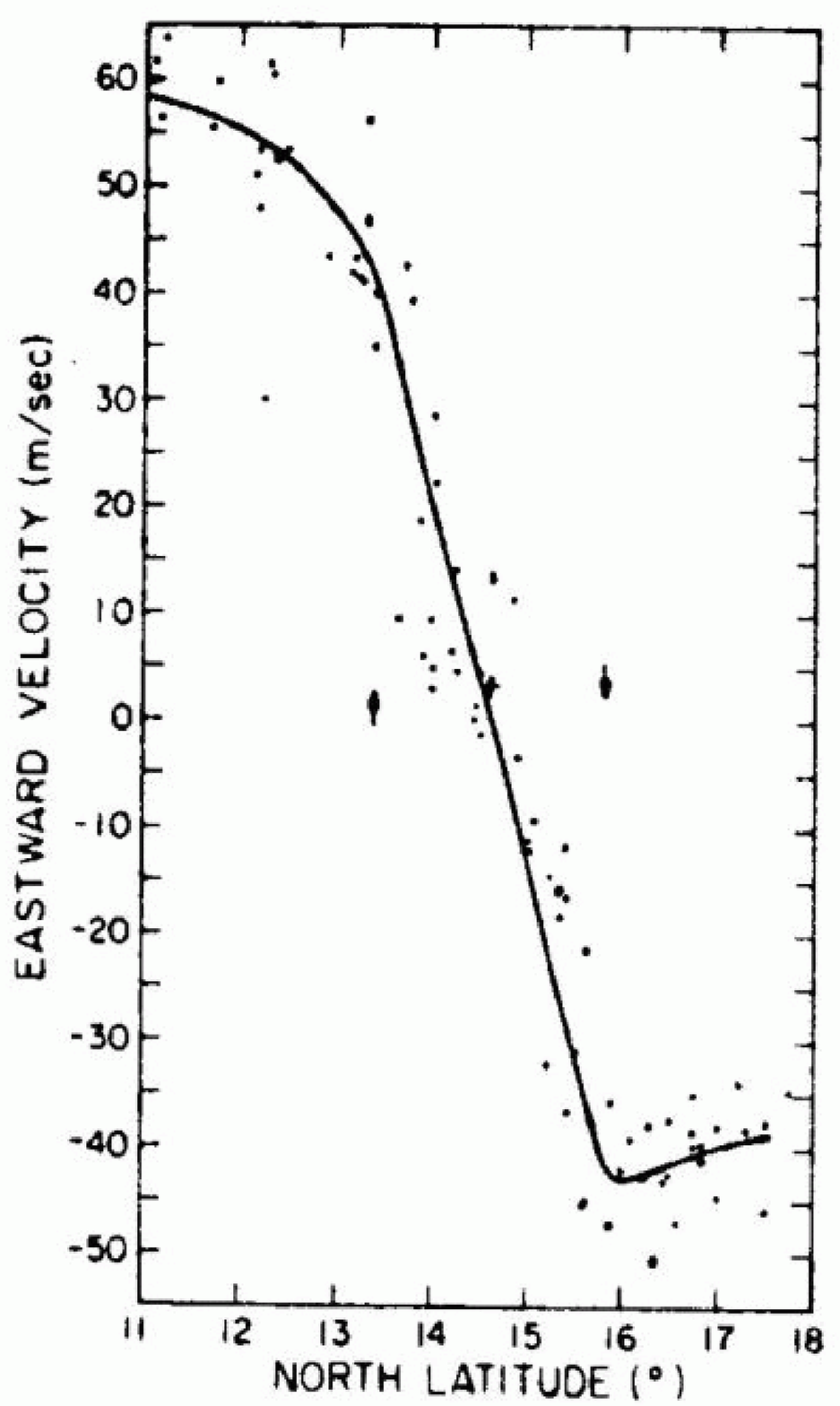}} &
\resizebox*{!}{0.4\textheight}{\includegraphics{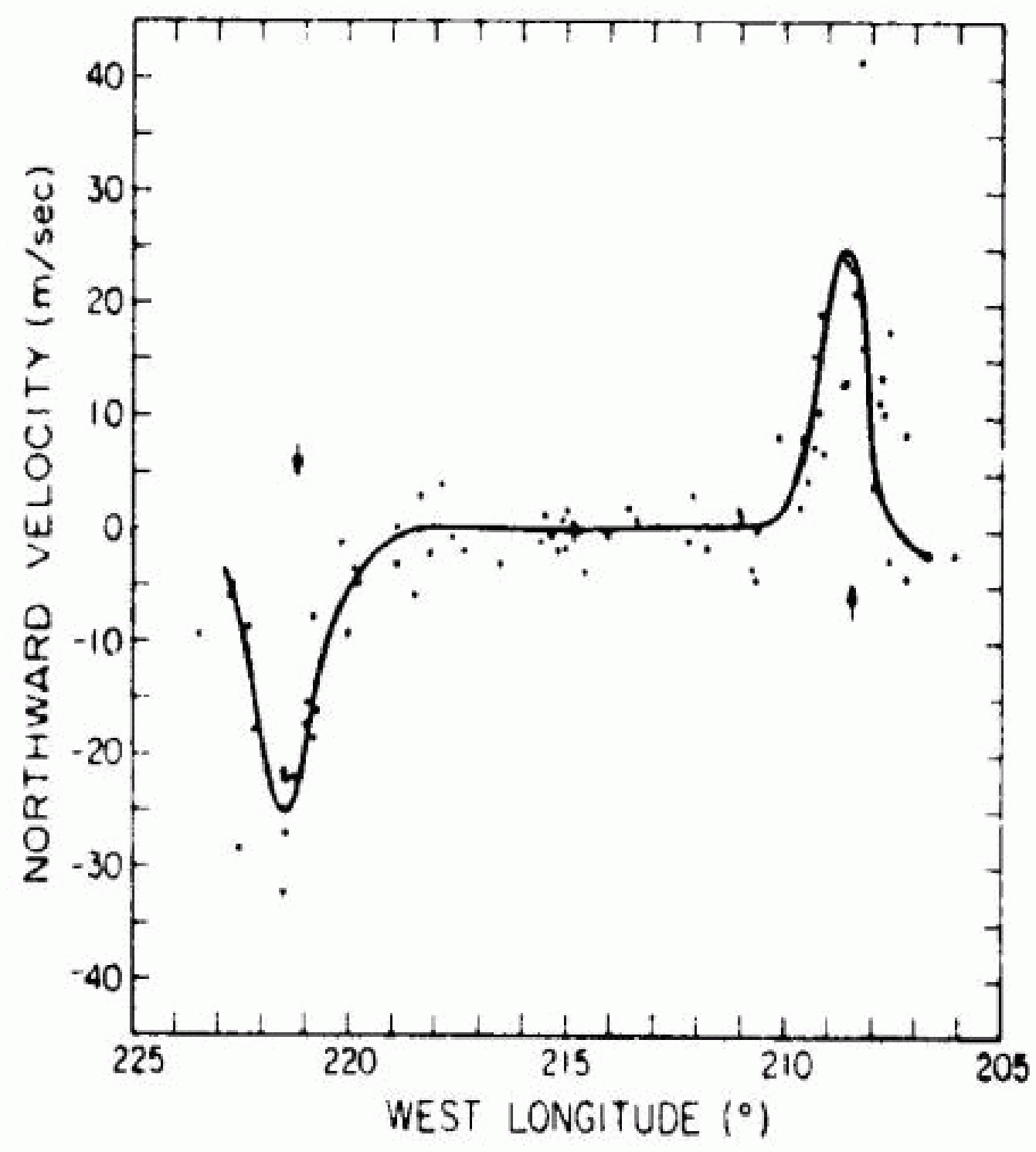}} \\
\end{tabular}\\
\begin{tabular}{cc}
\resizebox*{0.55\textwidth}{!}{\includegraphics{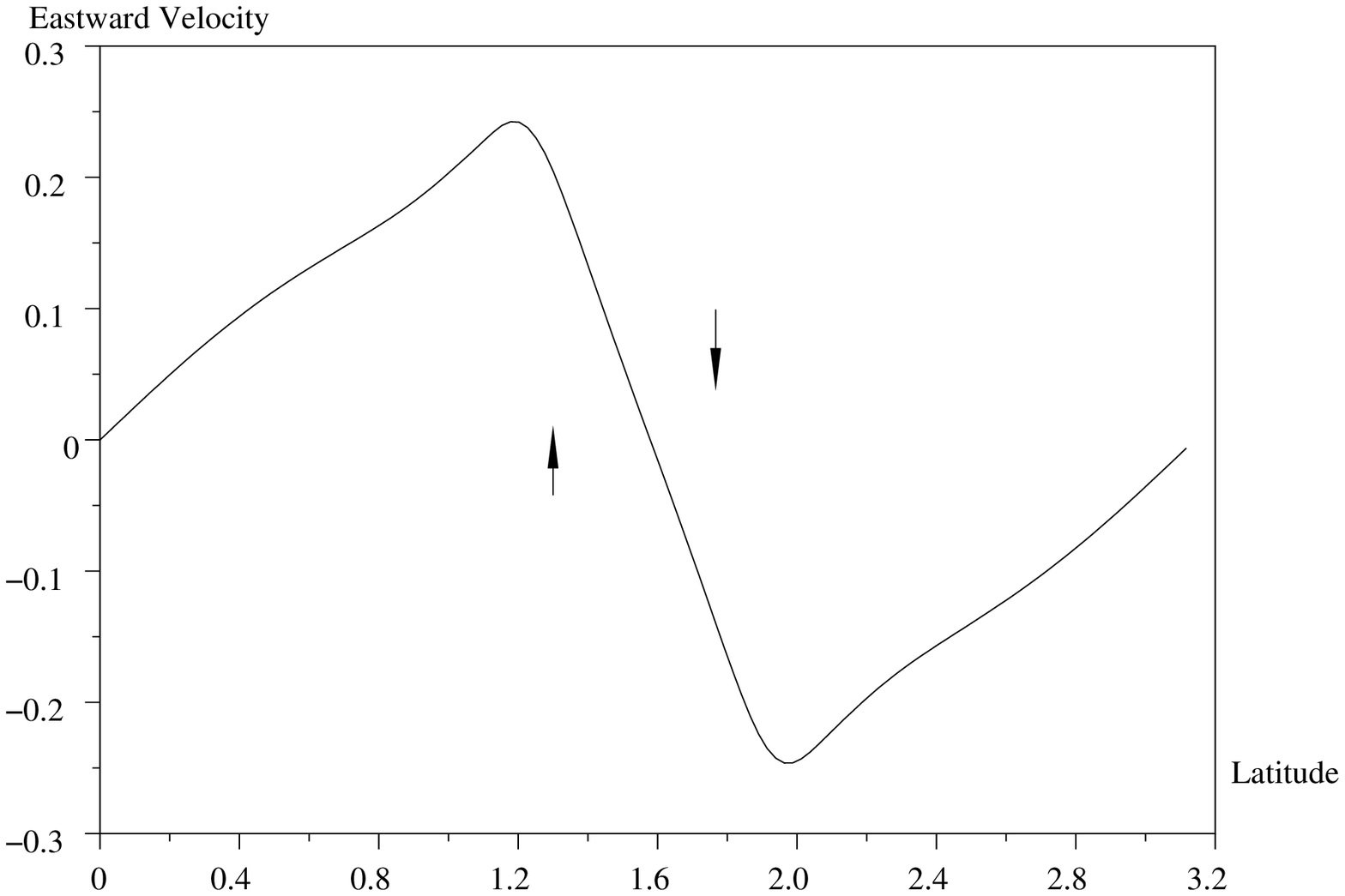}} &
\resizebox*{0.55\textwidth}{!}{\includegraphics{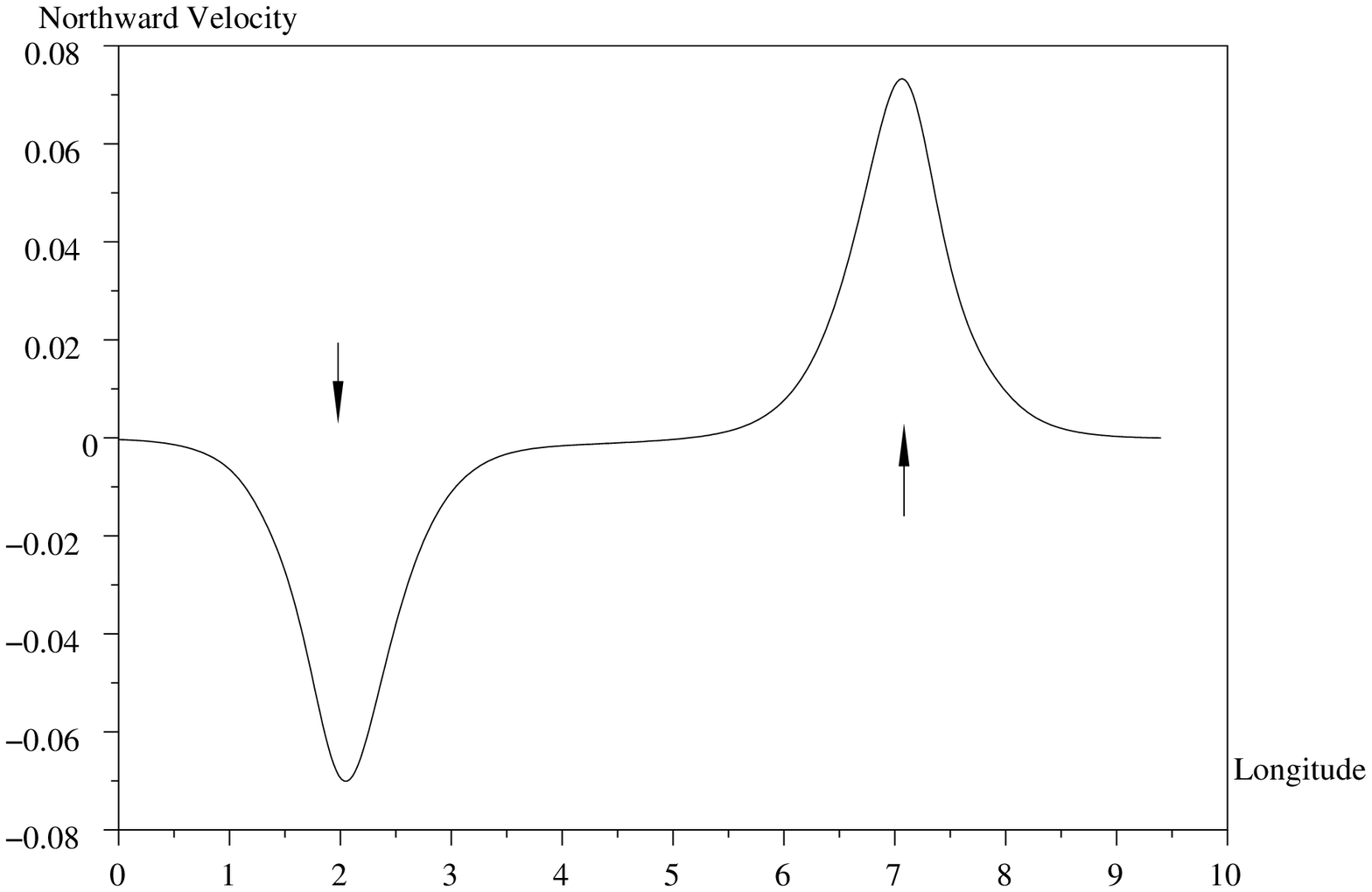}} \\
\end{tabular}\\
\end{tabular}\par}
\end{figure}

\end{document}